\documentclass[%
reprint,
superscriptaddress,
nofootinbib,
 amsmath,amssymb,
aps,
prx,
floatfix,
]{revtex4-2}

\usepackage{graphicx}
\usepackage{dcolumn}
\usepackage{bm}
\usepackage[hypertexnames=false]{hyperref}
\usepackage[all]{hypcap}    
\usepackage{siunitx}
\usepackage{braket}
\usepackage{rotating}
\usepackage{algorithm} 
\usepackage{algpseudocode} 
\usepackage{color}
\usepackage{xfrac}
\usepackage{subcaption}
\usepackage{ragged2e}

\begin{document}
\title{Extendable optical phase synchronization of remote and independent quantum network nodes over deployed fibers}

\author{A. J. Stolk}
\affiliation{\vspace{0.5em}QuTech, Delft University of Technology, 2628 CJ Delft, The Netherlands}
\affiliation{\vspace{0.5em}Kavli Institute of Nanoscience, Delft University of Technology, 2628 CJ Delft, The Netherlands}

\author{J. J. B. Biemond}
\affiliation{\vspace{0.5em}QuTech, Delft University of Technology, 2628 CJ Delft, The Netherlands}
\affiliation{\vspace{0.5em}Netherlands Organisation for Applied Scientific Research (TNO), P.O. Box 155, 2600 AD Delft, The Netherlands}

\author{K. L. van der Enden}
\affiliation{\vspace{0.5em}QuTech, Delft University of Technology, 2628 CJ Delft, The Netherlands}
\affiliation{\vspace{0.5em}Kavli Institute of Nanoscience, Delft University of Technology, 2628 CJ Delft, The Netherlands}

\author{L. van Dooren}
\affiliation{\vspace{0.5em}QuTech, Delft University of Technology, 2628 CJ Delft, The Netherlands}
\affiliation{\vspace{0.5em}Netherlands Organisation for Applied Scientific Research (TNO), P.O. Box 155, 2600 AD Delft, The Netherlands}

\author{E. J. van Zwet}
\affiliation{\vspace{0.5em}QuTech, Delft University of Technology, 2628 CJ Delft, The Netherlands}
\affiliation{\vspace{0.5em}Netherlands Organisation for Applied Scientific Research (TNO), P.O. Box 155, 2600 AD Delft, The Netherlands}

\author{R. Hanson}
\email{Correspondence to: R.Hanson@tudelft.nl}
\affiliation{\vspace{0.5em}QuTech, Delft University of Technology, 2628 CJ Delft, The Netherlands}
\affiliation{\vspace{0.5em}Kavli Institute of Nanoscience, Delft University of Technology, 2628 CJ Delft, The Netherlands}

\date{\today}
             
\begin{abstract}
Entanglement generation between remote qubit systems is the central tasks for quantum communication. Future quantum networks will have to be compatible with low-loss telecom bands and operate with large separation between qubit nodes. Single-click heralding schemes can be used to increase entanglement rates at the cost of needing an optically phase-synchronized architecture. In this paper we present such a phase synchronization scheme for a metropolitan quantum network, operating in the low-loss telecom L-band. To overcome various challenges such as communication delays and optical power limitations, the scheme consists of multiple tasks that are individually stabilized. We characterize each task, identify the main noise sources, motivate the design choices and describe the synchronization schemes. The performance of each of the tasks is quantified by a transfer-function measurement that investigates the frequency response and feedback bandwidth. Finally we investigate the resulting optical phase stability of the fully deployed system over a continuous period of 10 hours, reporting a short-term stability standard deviation of $ \sigma \approx30\deg$ and a long-term stability of the average optical phase to within a few degrees. The scheme presented served as a key enabling technology for an NV-center based metropolitan quantum link. This scheme is of interest for other quantum network platforms that benefit from an extendable and telecom compatible phase synchronization solution.
\end{abstract}

\maketitle

\section{Introduction Optical phase in Quantum networks }
\begin{figure}
\includegraphics[width = 0.9\columnwidth]{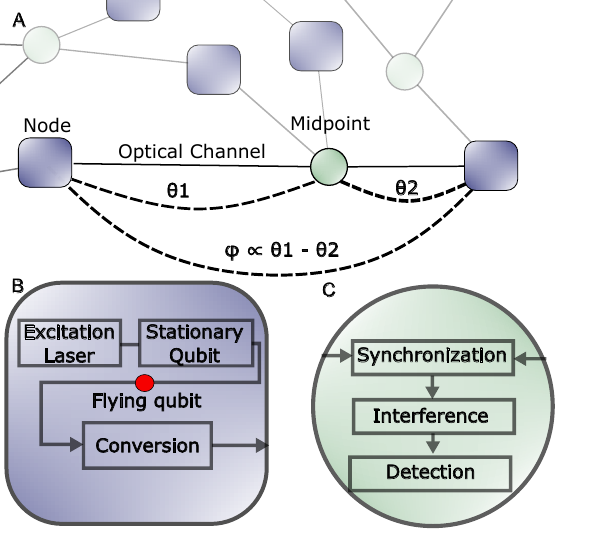}
\caption{\textbf{Schematic overview of elements in a Quantum Network.}\label{fig:Figure1} \textbf{A)} Many nodes can be connected by sending flying qubits over optical channels towards a central midpoint. The optical path from excitation in the node, to detection in the midpoint, gives rise to an optical phase, $\theta$. When using the Fock-state encoding of the flying qubit for entanglement generation, the resulting entangled state phase $\phi$ is proportional to the difference between the optical phases. B) Elements needed to generate entanglement between stationary qubits, via the exchange of flying qubits. If the flying qubits are photons not compatible with deployed infrastructure, frequency conversion can be used to significantly lower the propagation losses. C) Upon arrival in the midpoint, the phase of the incoming photonic qubits needs to be stabilized, after which they can be interfered and subsequently detected. The outcome of this detection heralds entanglement between the stationary qubits in the nodes.}
\end{figure}

Quantum networks~\cite{kimble_quantum_2008} hold the promise to revolutionize the way people exchange information. A central task of a quantum network is the generation of entanglement between (end-)nodes, in which the entanglement can be stored, manipulated and processed~\cite{wehner_quantum_2018}. In general, protocols to generate entanglement between stationary qubits involve the emission, transmission and joint measurement of flying qubits, encoded in a photon state, see Fig.~\ref{fig:Figure1}a. A common configuration envisioned for a future large scale quantum internet is the combination of nodes, which house the stationary qubits, and midpoints, where the flying qubit from different nodes are sent to. These nodes are connected via deployed telecom fiber to midpoints, that provide synchronization, interference and detection tasks, see Fig.~\ref{fig:Figure1}b. 

Photon loss in the connecting fibers decreases the rate at which entanglement between the nodes is generated, with the exact scaling depending on the photonic qubit encoding used. For example, for frequency encoding~\cite{moehring_entanglement_2007}, polarization encoding~\cite{ritter_elementary_2012,hofmann_heralded_2012, stephenson_high-rate_2020,Daiss2021,KrutyanskiyEntanglement2023}, and time-bin encoding~\cite{Barrett_Kok_2005,Hensen_bell_2015,Narla_Shankar_Hatridge_Leghtas_Sliwa_Zalys_Geller_Mundhada_Pfaff_Frunzio_Schoelkopf_etal._2016,Knaut_SiV_35kmBoston_2023}, the entangling rate scales linearly with the photon transmission $\eta$ between the nodes. 
In contrast, for photonic encoding using number states (0 or 1 photon)~\cite{Cabrillo_Cirac_García-Fernández_Zoller_1999, Bose_Knight_Plenio_Vedral_1999} the rate scales favourably with $\sqrt{\eta}$, yielding significantly higher rates in the typical scenario of substantial photon loss (i.e. for $\eta\ll1$).
However, with this encoding the resulting entangled state phase $\phi$ is proportional to the optical phase difference between the two paths to the midpoint ($\theta_1$ and $\theta_2$, see Fig.~\ref{fig:Figure1}a)), leading to the additional experimental requirement that this optical phase difference at the time of interference in the midpoint needs to be known \footnote{For time-bin encoding the entangled state phase is dependent on the optical phase difference picked up between the two time bins. Thus, optical phase stability is required only on this timescale which is typically below 500ns~\cite{bernien_heralded_2013,Hensen_bell_2015,Knaut_SiV_35kmBoston_2023}}.

If the phase deviates from the chosen setpoint by $\delta\phi$, the maximum fidelity achievable is given by $F(\delta\phi) = \frac{1}{2}(1+\cos(\delta\phi))$. In a larger network, the single photons are originating from multiple sources at distant locations, where the phase is affected at many length- and time-scales. Any attempt to stabilize this phase needs to take into account unwanted light from either conversion or the synchronization methods used, that degrade the fidelity of the entangled state. This puts stringent requirements on any classical stabilization light co-propagating with the quantum channel.

Entanglement generation using photonic number state encoding, hereafter called single-click protocol, has been demonstrated in atoms~\cite{Slodička_Hétet_Röck_Schindler_Hennrich_Blatt_2013}, (hole-)spins in semi-conductors~\cite{Delteil_Sun_Gao_Togan_Faelt_Imamoğlu_2016,stockill_phase-tuned_2017}, ensemble-based quantum memories~\cite{Lago-Rivera_telecommemories_2021, LiuTelecomMemories2024}, and between single rare-earth ions in cavities~\cite{RuskucREIentanglement2024}.
Specifically, the implementation of this protocol on the Nitrogen Vacancy (NV-) center in diamond~\cite{Humphreys_deterministicdelivery_2018} resulted in orders of magnitude faster entanglement generation than the preceding experiments on the same platform that used time-bin encoding~\cite{bernien_heralded_2013,Pfaff2014,Hensen_bell_2015}. This allowed for the extension to a multi-node network on which distributed quantum protocols can be realized~\cite{pompili_multinodenetwork_2021, Hermans_Pompili_Beukers_Baier_Borregaard_Hanson_2022}. So far these experiments have not dealt with the additional requirements of large separation between nodes or telecom compatibility, allowing them to simplify their design. Our approach forms an essential part of a metropolitan quantum link realized in the Netherlands, where solid-state entanglement is generated between two NV-based quantum nodes using \SI{25}{km} of deployed telecom fiber\cite{preprint}. 

A key challenge for the implementation of the single-click protocol for entanglement generation over large distances is the strict requirement on the optical phase stability. Large physical separation between the end nodes adds even more complexity to the system, as fast fluctuation of the optical phase can only be synchronized by using high-bandwidth feedback, where the propagation-delay of information exchange between locations becomes potentially problematic. Furthermore, directly sharing of optical phase references between the nodes beforehand becomes challenging when the distance between the nodes becomes larger and the excitation lasers are in the visible wavelengths. The duration of this stability is demanded for the full duration of the time it takes to analyze (or in the future, end-user protocol runtime) the entangled state that is being generated. A robust, extendable and highly synchronized solution is therefore a key enabler for future quantum networks at scale.

\captionsetup[figure]{justification=raggedright}

In this work we propose, implement and verify an optical phase synchronization scheme between remote and independent quantum nodes operating in the telecom band. The structure of the paper is as follows. In the following section we provide the broader scope in which our phase synchronization scheme is developed and the challenges it is aiming to solve. In the section after that, we provide a full system overview and provide information on the various noise sources that occur in the system. We describe the division of the full system into smaller synchronization tasks, which are subsequently individually described, where the control-layout, noise spectrum and feedback performance are discussed. We then verify the synchronization of the full system at work using measurements of the optical phase between the nodes deployed in the network. Finally we conclude by highlighting the benefits of our scheme, as well as give an outlook on broader applications and further improvements.

\section{Phase-stabilization for entanglement generation using NV-center based processors and telecom fibers}

\begin{figure*}[hbt!]
\includegraphics[width=0.7\linewidth]{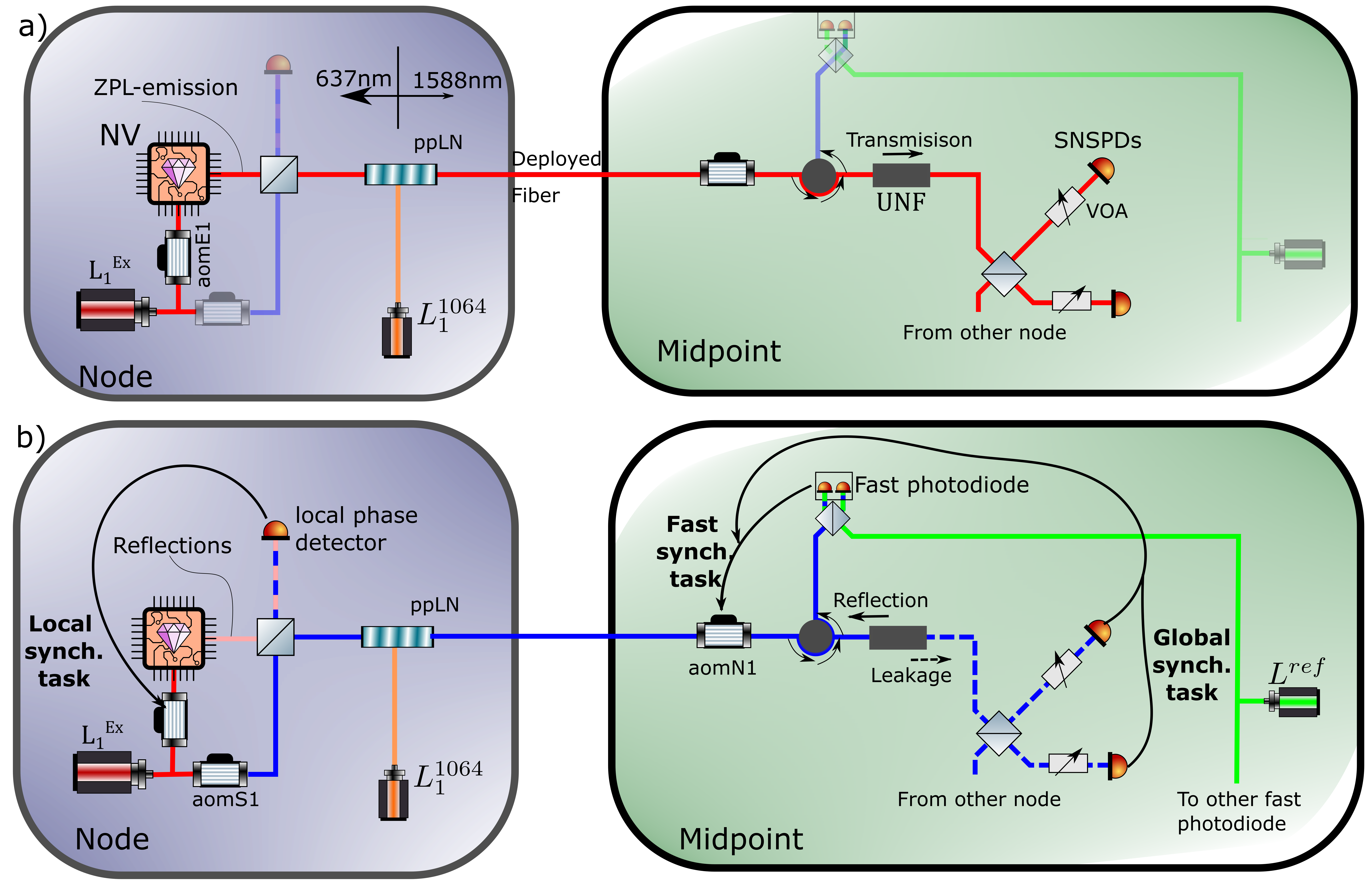}
\caption{\label{fig:opticsphaselock}\textbf{Schematic of optical lay-out with sources, detectors and optical phase actuators. a)} Optical paths for entanglement generation. At the node light from the excitation laser (red) excites the NV center, yielding single photons at the ZPL wavelength (in red) through spontaneous emission which are converted and routed to the midpoint. There the single photons are passed through an AOM and an ultra-narrow FBG (UNF) for spectral filtering. Overlapping the incoming photon with the photon from the other node on a central beamsplitter, they are measured using SNSPDs. As described in the main text, the relative phase at this beamsplitter is crucial in the entanglement generation.
\textbf{b)} The lay-out of the phase synchronization schemes to achieve the synchronized phase at the beamsplitter in the midpoint. The laser used for excitation of the NV-center is split and offset in frequency to generate stabiliation light (in blue). In addition reflections of the excitation light (shown pink) coming from the diamond surface, are generated which guided to the `local phase detector', interfering with part of the stabilization light. This error signal is used to feedback the local synchronization task. The stabilization light is also converted and sent to the midpoint, following the same optical path as the single-photons. There the stabilization light co-propagates with the single photons until it hits the UNF in the rejection band, where most of the stabilization is reflected and routed towards the fast photo-diode, ultimately interfering at the `fast detector' with light from a reference laser (green). This error signal is used to synchronize the fast task. Leakage stabilization light passes through the FBG (blue dotted line) and reaches the central beamsplitter at the midpoint and the SNSPDs. Similarly, light from Node 2 arrives at the midpoint to arrive at the same beamsplitter, interfering with the light from Node 1. This beat can be measured by the SNSPDs, and the error signal is used to stabilize the global synchronization task.}
\end{figure*}

The NV center is an optically active defect in a diamond lattice, of which the electronic spin state forms the basis of many previous demonstrations of solid-state entanglement generation in a network~\cite{bernien_heralded_2013, Hensen_bell_2015, pompili_multinodenetwork_2021}. The entanglement generation can be briefly described in three steps: (1) the generation of single photons by resonant excitation and spontaneous emission, (2) the collection, possible frequency conversion, and propagation towards a central beamsplitter, and (3) a measurement of a single photon. The schematic in the top of Fig.~\ref{fig:opticsphaselock} shows the important optical components needed to perform the entanglement generation.

Single-photons are generated by carving short (\SI{1.5}{\nano\second} FWHM) optical pulses from a tap-off from a continuous laser. This excitation light is routed to the NV-center housed in a \SI{4}{\kelvin} cryostat via in-fiber and free-space optics. Spin-selective, resonant excitation followed by spontaneous emission, generates single photons emitted by the NV-center (Fig. \ref{fig:opticsphaselock}a, left). Around $3$\% of the photon-emission is not accompanied by a phonon emission, the so called Zero-Phonon line (ZPL). These photons are coherent with the laser field used for excitation, and are entangled with the spin-state of the NV-center. The ZPL photons are collected into a single-mode fiber, and guided to a Quantum Frequency Converter (QFC), similar to previous work~\cite{Dréau_Tchebotareva_Mahdaoui_Bonato_Hanson_2018,Geus_Elsen_Nyga_Stolk_vanderEnden_vanZwet_Haefner_Hanson_Jungbluth_2023}, that maintains the entanglement between the photon and the NV spin~\cite{Tchebotareva_TelecomSpinPhotonEntanglement_2019}. 

In the QFC process, the single-photons are mixed with a high-power pump (\SI{1064}{\nano\meter}) inside a non-linear medium, which is phase-matched for a Difference Frequency Generation process (Fig.\ref{fig:opticsphaselock}a). This converts the single photons from the original \SI{637}{\nano\meter} to \SI{1588}{\nano\meter} in the telecom L-band. This process is crucial to reduce propagation losses, and also serves as a method to remove any frequency difference between the different NV-centers used for entanglement generation. For more details see previous experiments ~\cite{Stolk_VanDerEnden_Roehsner_Teepe_Faes_Bradley_Cadot_VanRantwijk_TeRaa_Hagen_etal_2022}.

After the frequency conversion, the single-photons propagate over the deployed fibers towards a central location called the midpoint, shown in Fig.~\ref{fig:opticsphaselock}a on the right. There the incoming photons are spectrally filtered via transmission through an Ultra-narrow Filter (UNF, FWHM \SI{50}{\mega\hertz}), and guided to an in-fiber beamsplitter, where the modes of the two nodes interfere. At this point, the relative optical phase is crucial for the entanglement generation when using the single click protocol. The resulting modes behind the beamsplitter are measured using Superconducting Nanowire Single Photon Detectors (SNSPDs), that perform a photonic Bell-state measurement on the combined photon mode. The outcome of this measurement heralds the spin-state of the NV-centers in an entangled state.

For this configuration of a qubit platform we highlight a few specific challenges of stabilizing the optical phase.

Firstly, the free-space and in-fiber optics is sensitive to resonances when placed on an optical table together with a closed-cycle cryostat inducing vibrations, showing up as phase noise on the single photon field. Depending on the mechanical frequencies and stability, a moderate feedback bandwidth is needed to get rid of these fluctuations. Furthermore, careful design of the optics should be done to limit these mechanical vibrations, to reduce the phase-noise that is present due to this effect.

Second, the optical path that requires phase-synchronization is inherently optically connected with a two-level quantum system on one end, and sensitive single-photon detectors on the other end. This constrains the optical powers that can be used for the phase synchronization, and makes the design of the whole system more complex. For instance, unwanted reflections at fiber to fiber connectors or free-space to matter interfaces can lead to crosstalk between the synchronization tasks and entanglement generation. The careful balancing of optical powers and illumination times of light that is incident on the NV-center, as well as the extra shielding of the SNSPDs with Variable Optical Attenuators (VOAs) are additions to allow for a more stable optical phase without reducing the coherence of the NV-center or blinding of the SNSPDs. 

Third, when using optical fibers over large distance, the thermal expansion can introduce variations that expand the fiber in the same direction for days. Because we only stabilize the relative phase, large drift in fiber length introduces many phase-slips of $2\pi$, which results in a non-negligible difference between the stabilized phase and optical phase of interest, see \ref{subsec:phaseslipsub}. This is due to the fact that the stabilization light propagates over the long fiber with slightly different frequency than the photons with which we generate entanglement. In our case, when using an offset of \SI{400}{\mega\hertz}, residual phase error $\Delta\theta$ is $\Delta\theta = M\ 360\ (\frac{f_{NV} - f_{stab}}{f_{stab}}) = M\ 7.6\times10^{-4}$ degrees, where $M$ is the number of phaseslips. For deployed fibers of kilometers long, many centimeters of expansion/contraction occurs on the timescale of days, corresponding to millions of phaseslips. This makes it significant for continuous operation of entanglement generating networks over large fiber networks.

Fourth, propagating high power optical pulses over long fibers can create a significant background at the single photon level both due to double Rayleigh scattering~\cite{Chen2020DRS}, and via interfaces such as connectors and splices. This puts limits on the shot-noise limited feedback bandwidth one can achieve using this classical light, without inducing more background photons in the SNSPDs.

In the following section we discuss our approaches to tackle all these challenges, given the boundary conditions as discussed in the previous sections. We identify three subtasks that can be synchronized independently, resulting in the synchronization of the relative phase of the single-photon fields between two NV-center nodes.

\begin{table*}[]
    \centering
    \begin{tabular}{|p{0.2\linewidth}|p{0.25\linewidth}|p{.29\linewidth}|p{.1\linewidth}|p{.15\linewidth}|}
    \hline
        \textbf{Synchronization task} &\textbf{Light sources generating error signal and used actuator}&\textbf{Distortions and dominant frequency range}&\textbf{Heterodyne frequency}&\textbf{Achieved \SI{-3}{\deci\bel} feedback bandwidth} \\\hline
          Local synchronization at the node & reflected excitation light from Quantum Device and stabilization light measured on photodiode. Feedback implemented on AOM of excitation light. &
          \begin{minipage}[t]{.99\linewidth}    
            \begin{itemize}
              \item mechanical vibrations of objective lens w.r.t sample, \SI{10}{\hertz}-\SI{5}{\kilo\hertz}
              \item thermal drift of free space/fiber optics $<$\SI{1}{\hertz}
            \end{itemize}
          \end{minipage}
          &\SI{400}{\mega\hertz} 
          $\pm$\SI{750}{\hertz}& \SI{3}{\kilo\hertz}
          \\\hline
          Fast synchronization at the midpoint  
          &Stabilization light of selected node and reference light, measured on fast detector. Feedback via AOM acting on both the stabilization light and single photons. Actuator desaturation is done via the pumplaser of QFC.
          &
                    \begin{minipage}[t]{.99\linewidth}    
            \begin{itemize}
              \item thermal effects on deployed fiber $<$\SI{10}{\hertz}
              \item mechanical vibrations of deployed fiber \SI{1}{\hertz}-\SI{5}{\kilo\hertz}
              \item phase noise in excitation laser, pumplaser and reference laser \SI{1}{\kilo\hertz}-\SI{50}{\kilo\hertz}
              \end{itemize}
          \end{minipage}
          &\SI{215}{\mega\hertz} +0/-\SI{1500}{\hertz}
          &\SI{224}{\kilo\hertz}
          \\\hline
          Global synchronization at the midpoint 
          &Stabilization light of node 1 and 2 measured on SNSPDs. Compensation through the setpoint of the fast synchronization loop from node 1.
          &          \begin{minipage}[t]{.99\linewidth}    
            \begin{itemize}
              \item thermal drift at midpoint $<$\SI{10}{\hertz}
            \end{itemize}
          \end{minipage}
          &\SI{1500}{\hertz}
          &$\sim$\SI{50}{\hertz}
        \\\hline
    \end{tabular}

    \caption{Summary of synchronization tasks.}
    \label{tab:synctasks}
\end{table*}

\captionsetup[figure]{justification=raggedright}
\begin{figure*}
\includegraphics[width=1.5\columnwidth]{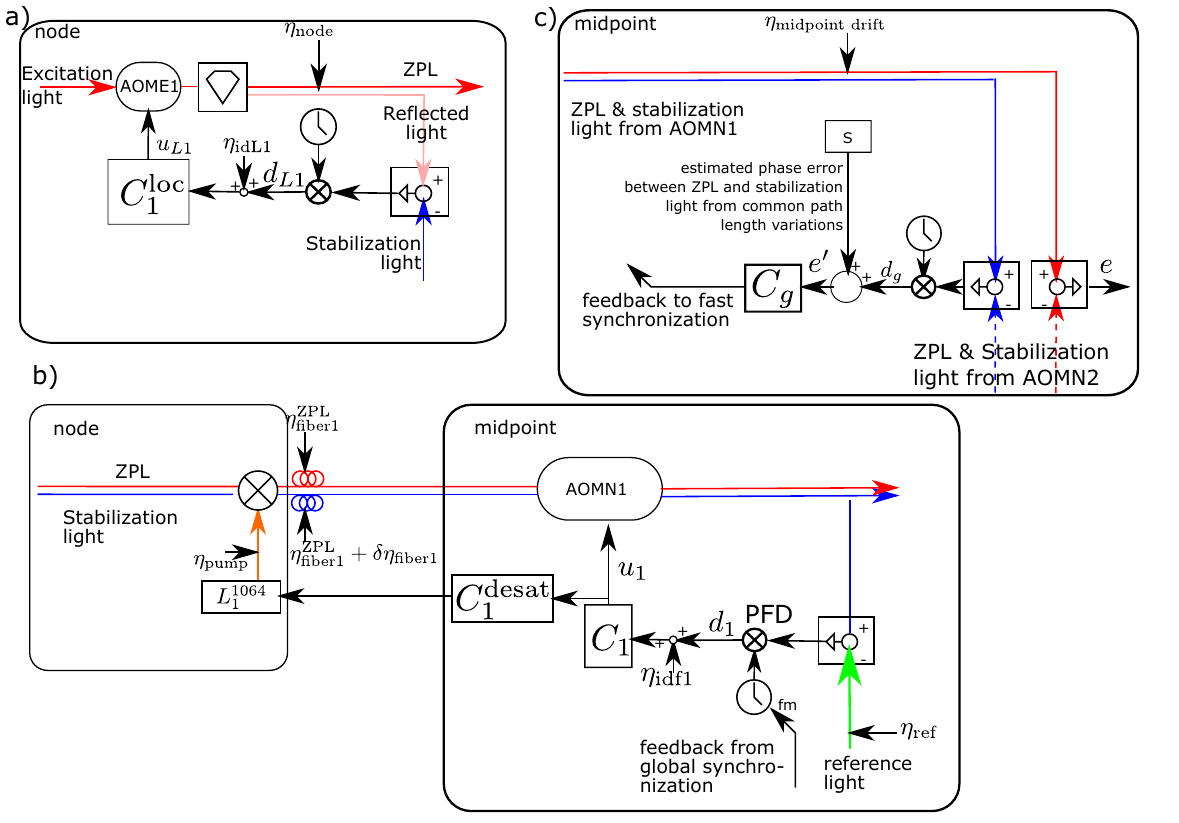}
\caption{\label{fig:controlscheme}\textbf{Control scheme including main distortions. a) Local synchronization.} At the nodes, light exciting the NV is modulated using AOME1, where the controller $C_{1}^\text{loc}$ applies a frequency modulation of the sine generator driving the AOM. This modulation depends on the heterodyne measurement of the beat between reflected and stabilization light at a balanced detector. Main noise sources are the phase noise of the excitation laser and mechanical vibrations and drift that are not shared by the stabilization light and the ZPL light ($\eta_\text{node}$). 
\textbf{b) Fast synchronization.} Subsequently, ZPL and stabilization light are frequency-converted by mixing with a 1064nm-pumplaser, and this light travels from both nodes to the midpoint. The stabilization light is separated from the ZPL light and interferes on a balanced detector with the reference laser, where the demodulation is performed with a Phase Frequency Detector (PFD). The PFD output is fed to a synchronization controller $C_{1}$ which modulates the ZPL and stabilization light with aomN1. This fast controller suppresses mechanical distortions $\eta_\text{fiber}^\text{ZPL}$ shared between ZPL and stabilization light and phase noise from the excitation- and pumplaser, i.e. $\eta_\text{excitation}$ and $\eta_\text{pump}$. A varying setpoint to the fast synchronization task on node 1 is supplied by the global synchronization task.
\textbf{c) Global synchronization}. The  phase error between stabilization light at the SNSPDs is measured to suppress phase drifts $\eta_\text{midpoint drift}$ in the optical paths at the midpoint between reference laser and fast detectors, as well as between the fast detectors and the SNSPDs. This  heterodyne measurement uses the SNSPD count rates as error signal. The global controller $C$ updates the setpoint of the fast synchronization task of node 1. Since larger length variations will occur over the deployed fiber, a phase error will be introduced between ZPL and stabilization light, which is estimated using fiber-length measurements and compensated through $S$. Homodyne interference between ZPL light also is measured using the SNSPDs. The interference between ZPL light (red) is used in the entanglement generation process, and the error $e$  directly reduces the maximum fidelity thereof.}
\end{figure*}

\section{System overview and noise sources\label{sec:sysoverview}}
An overview of the optical layout that enables synchronization is shown in Fig.~\ref{fig:opticsphaselock}b. The strategy to divide the phase synchronization for NV-based networks in multiple subtasks was first divised and implemented in Ref.~\cite{pompili_multinodenetwork_2021}, which we extend in this work to suit the additional challenges of telecom operation and large node separation. In this section we identify those sub tasks, showing which optical paths are used where error-signals are generated and subsequently what actuator is used to achieve the synchronization. In Fig.~\ref{fig:controlscheme} we describe the synchronization subtasks in more detail, giving an overview of the individual components used.

To achieve the ultimate goal of synchronizing the ZPL photons coming from different nodes, an additional light field is added on both nodes, enabling optical phase synchronization at higher optical powers. This stabilization light is generated at both nodes from the same laser that also excites the NV, and is offset in frequency using a similar set of AOMs. This stabilization light is optically overlapped with light coming from the cryostat using a beamsplitter on the node, where the majority is transmitted and overlaps with the reflections coming from the diamond chip. This reflection is used to synchronize the ZPL light to the stabilization light via a local interferometer in the node, see Fig.~\ref{fig:opticsphaselock}b, left. This interferometer stabilization, called the local synchronization task, is discussed in Section~\ref{sec:locallock}, and effectively makes the stabilization light a good phase reference coherent with the ZPL light.
The part of the stabilization light that does not go to the local photodiode, propagates via the same fiber as the ZPL photons, via the QFC to the midpoint. 
Fast phase disturbances in the fiber and phase noise from the excitation and pump lasers are now present on both the stabilization light and ZPL light. The stabilization light is separated from the ZPL using a spectral filter due to the frequency offset of \SI{400}{\mega\hertz}. The stabilization light is synchronised to light from a telecom reference laser by means of frequency modulation by of an AOM driver. We do this by generating an error signal on a balanced photodetector, as shown in Fig. \ref{fig:opticsphaselock}b. This is called the fast synchronization lock and is described in more detail in Section~\ref{sec:fastmidpointlock}. 

Finally, the SNSPDs are used to assess the optical powers of the stabilization light to each detector, generating a single-photon detector based detection scheme for interference measurement. The error signal is generated using stabilization light that is leaking through the spectral filter at both arms, and meets on the central beampslitter connecting the two arms, see Fig.~\ref{fig:opticsphaselock}b. This measurement is inherently of low bandwidth, limited by the low powers used, and the SNSPDs' maximum count rate of \SI{1}{\mega\C\per\second}. Higher bandwidths would demand lower integration times, and introduce too high shotnoise. Based on this phase measurement, the setpoint of the fast synchronization task 1 is altered. This allows the synchronization of the stabilization light combining from the two nodes, closing the synchronization scheme. As length variations (caused e.g. by temperature variations) over the long deployed fiber will generate non-negligible phase variations between light at the ZPL and stabilization frequency, this length variation is measured and the frequency-induced phase variation is compensated. Using this compensation, synchronization between ZPL light from both nodes is achieved, cf.~Section~\ref{sec:globmidpointlock}.

Hence, 3 distinct synchronization tasks have been identified. Each of these synchronization tasks are performed using heterodyne phase measurements, and are described in more detail in the following three sections. An overview of the three tasks and their description, including noise sources, heterodyne frequency and feedback bandwidth is shown in Table~\ref{tab:synctasks}. A full theoretical description of the synchronization tasks, underlying assumptions and the conditions that must be met for them to function is given in the supplementary materials \ref{suppsec:theodescription}.

\section{Local synchronization at the node\label{sec:locallock}}
The local synchronization tasks consists of ensuring that the reflected excitation light from the Quantum Device is phase-synchronized to the stabilization light. In this manner, we also ensure phase synchronization between the ZPL photons and the stabilization light. Namely, the ZPL light emitted at the NV center is phase-synchronous with the excitation light as they run over the same optical path\footnote{There is a small optical path difference between the emitter location and light reflected off the diamond surface, which can be kept constant via spatial optimization of the microcsope objective.}. The reflected excitation light is separated from the ZPL light based on polarisation using a free space beamsplitter. The stabilization light enters the system at the other input, after which we use a set of birefringent $\alpha$-Bariumborate ($\alpha$BBO) crystals to maximize the complex overlap between the orthogonally polarized reflected beam and the stabilization light. These crystals can correct for static differences in tip-tilt and translation errors between the stabilization and reflected light. After this optimization, we project both beams in a common polarization mode using a polarizer, and measure a sufficient interference signal using a photodiode. After demodulation, an analogue proportional controller with roll-off filter is used to generate a frequency-modulating signal for the AOM driver of AOME1, see Fig.~\ref{fig:opticsphaselock}.

To assess the performance of this control loop, a \emph{linear} system identification experiment is performed for each of the synchronization tasks, as show in Fig.~\ref{fig:controlscheme}a. Within each synchronization task we can identify a `plant'. The plant is defined as the combination of the processing and actuation of the error signals (e.g. $d_{L1}$ to $u_{L1}$). It consists of the actions of the frequency-modulation of a digital clock, amplification to drive the AOM, interference detection and demodulation of the electrical signal. While the frequency shifting behaviour of the AOM and subsequent demodulation via an analogue diode-based double balanced mixer are in practice nonlinear, we assume a linear input-output behaviour of the plant, which matches our observations.

To identify this subsystem\footnote{In contrast to the final operation, the stabilization light is constantly available, i.e. the \SI{2.5}{\micro\second} `dark' periods every \SI{10}{\micro\second} period are avoided for the identification experiments.} in series with the controller we use a commonly used technique of injecting an additional but known broadband noise signal on the actuator ($\eta_\text{idL1}$ and $\eta_\text{idL1}$ in Fig.~\ref{fig:controlscheme}a. By measuring the resulting output with the feedback on and off and comparing them we can calculate important performance parameters such as the open-loop transfer function of the plant. We can also record the spectral densities of the residual phase noise. For more information see the supplementary information.

\captionsetup[figure]{justification=raggedright}
\begin{figure}[!h]
\includegraphics[width=0.9\columnwidth]{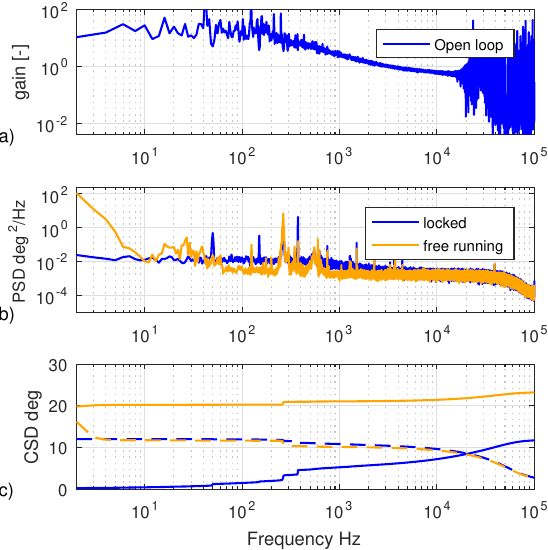}
\caption{\label{fig:localmerged}\textbf{Performance of local synchronization task}. a) Identified linear dynamical relation between control output $u_{L1}$ and demodulation output $d_{L1}$ (demodulated around DC as used as control input), called plant. b) Power Spectral Density and c) Cumulative (solid) and Inverse (dotted) Cumulative Spectral Density of residual phase error with (blue) and without (orange) feedback, measured without additional noise injection.}
\end{figure}

We show the result of this analysis for the local synchronization task in Fig.~\ref{fig:localmerged}. It shows that a bandwidth of $\sim$\SI{3}{\kilo\hertz} is obtained, where the gain reaches a value of $\sim0.5$ in Fig.~\ref{fig:localmerged}a. The free-running Power Spectral Density (PSD) shown in Fig.~\ref{fig:localmerged}b (orange) has the features of slow drifts below \SI{10}{\hertz}, as well as mechanical resonances between \SIrange{100}{1000}{\hertz}. By integrating the PSD when the controller is enabled (Fig.~\ref{fig:localmerged}b, blue), we reach a cumulative phase error of \SI{12}{\degree} RMS, as shown in Fig.~\ref{fig:localmerged}c.

\section{Fast synchronization at the midpoint\label{sec:fastmidpointlock}}
Both fast synchronization loops at the midpoint have the goal to synchronize the stabilization light of the nodes to light of the same reference laser which is located in the midpoint, see Section~\ref{sec:sysoverview}b and Fig.~\ref{fig:controlscheme}b, and have the same design for both arms coming from Node 1 and Node 2, and we discuss the workings in the context of Node 1. After conversion and subsequent arrival in the midpoint, stabilization light of the node and reference light interfere and is measured with a balanced photodetector, demodulated using a phase frequency detector (PFD) and then passed to an analog controller $C1$ of proportional-integral type. The output of this controller is then used for frequency modulation of the AOM AOMN1. To limit the frequency-modulation range of this AOM to avoid reduced optical transmission, de-saturation is required, which is performed via additional feedback to the pump laser frequency back at the node. This control signal is sent over an User Datagram Protocol (UDP) connection (update frequency \SI{500}{\hertz}) from the midpoint to the node. 

\begin{figure}[!h]
\includegraphics[width=0.9\columnwidth]{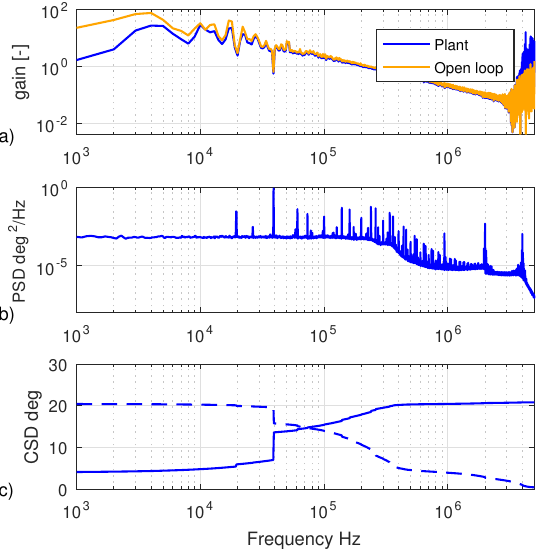}
\caption{\label{fig:fast1merged}\textbf{Performance of fast synchronization task.} a) Identified linear dynamical relation between control output $u_1$ and demodulation output $d_1$ (demodulated around DC as used as control input), called plant, and between control input and demodulation output, i.e. the open-loop behaviour. 
b) Power Spectral Density and c) Cumulative and Inverse Cumulative Spectral Density of residual phase error.}
\end{figure}

To identify the synchronization performance, we used the same techniques as outlined for the local synchronization task. The resulting open-loop transfer function is identified and shown in Fig.~\ref{fig:fast1merged}a. It shows that a \SI{220}{\kilo\hertz} control bandwidth is achieved. By integrating the `locked' PSD of the residual phase noise (Fig.~\ref{fig:fast1merged}b), we find the cumulative phase error of \SI{21}{\degree} RMS, as shown in Fig.~\ref{fig:fast1merged}c. The achieved bandwidth is realized by the short distance between error-signal generation and actuation, and fast servo control for error signal processing. Spectral data of the recorded signals is shown as a supplementary figure at the top of Fig.~\ref{fig:psdscombined}a.  

\captionsetup[figure]{justification=raggedright}

\section{Global synchronization\label{sec:globmidpointlock}}
The global synchronizing feedback is required to suppress low-frequent phase drifts in the optical paths at the midpoint between reference laser and fast detectors, as well as between the fast detectors and the SNSPDs, schematically depicted in Fig.~\ref{fig:controlscheme}c. A heterodyne phase error measurement is obtained from the SNSPD measurement by converting the single-photon count rate to well-defined pulses using an analog pulse stretcher, subtracting both pulse signals and lowpass-filtering the result. Employing a small frequency offset between the stabilization light between the nodes, we can detect a heterodyne beat at \SI{1500}{\hertz}, still below the relatively low bandwidth of the error signal generated with the SNSPDs, while making the measurement insensitive to power fluctuations (see Table~\ref{tab:synctasks} for details).
The error signal is processed and fed back by changing the setpoint of one of the fast synchronization scheme by an amount determined by controller C', which affects the phase shift introduce by AOMN1, cf.~\cite{patent_sync}. A proportional controller is used in this setup.

We compensate the expected phase differences between stabilization and ZPL light over the deployed fibers (i.e., $\delta \eta_\text{fiber1}$ and $\delta \eta_\text{fiber2}$, which will occur due to the optical frequency mismatch, combined with length variations over these fibers (see also \ref{suppsec:defassumptions}). Exploiting the measurement of these length variations using roundtrip-time measurements, estimates  $\tilde\delta \eta_\text{fiber1}$ and $\tilde\delta \eta_\text{fiber2}$ are obtained in subsystem S in~Fig.~\ref{fig:controlscheme}, cf.~\cite{patent_chromaticdiff}.

\begin{figure}
\includegraphics[width=0.9\columnwidth]{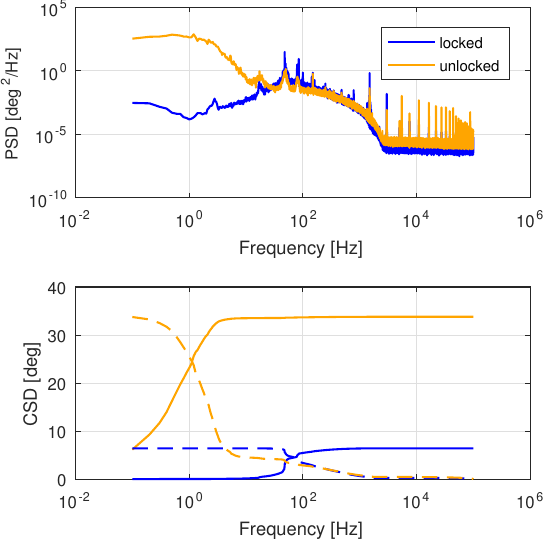}
\caption{\label{fig:globalpsd}\textbf{Performance of global phase synchronization}. Power Spectral Density (top) and Cumulative and Inverse Cumulative Spectral Density (bottom) of residual phase error.
}
\end{figure}

The performance of the global synchronization is illustrated in Fig.~\ref{fig:globalpsd}, where the spectrum with and without this feedback is shown. The dominant noise-sources are below \SI{10}{\hertz}, as seen in the unlocked PSD of the phase noise in Fig.~\ref{fig:globalpsd}. This is because all high-frequency noise is removed by the fast synchronization task, and only slow drifts remain. This controller yields a cumulative phase error of \SI{8}{\degree} RMS, as shown in Fig.~\ref{fig:globalpsd}c.

\captionsetup[figure]{justification=raggedright}

\section{Synchronization result on relative phase between remote nodes. \label{sec:sync2qi}}
We now measure the performance of the complete system when all the synchronization tasks are active. We evaluate the system by measuring the phase error $e$ between the excitation lasers of the both nodes, see Fig.~\ref{fig:controlscheme}c. The system is deployed at three locations in the Netherlands, where the nodes are separated by $\approx$\SI{10}{\kilo\meter}, connected to the midpoint by \SI{10}{\kilo\meter} and \SI{15}{\kilo\meter} of fiber between the cities The Hague and Delft respectively.

\captionsetup[figure]{justification = raggedright}
\begin{figure*}[h]
\includegraphics[width=0.75\textwidth]{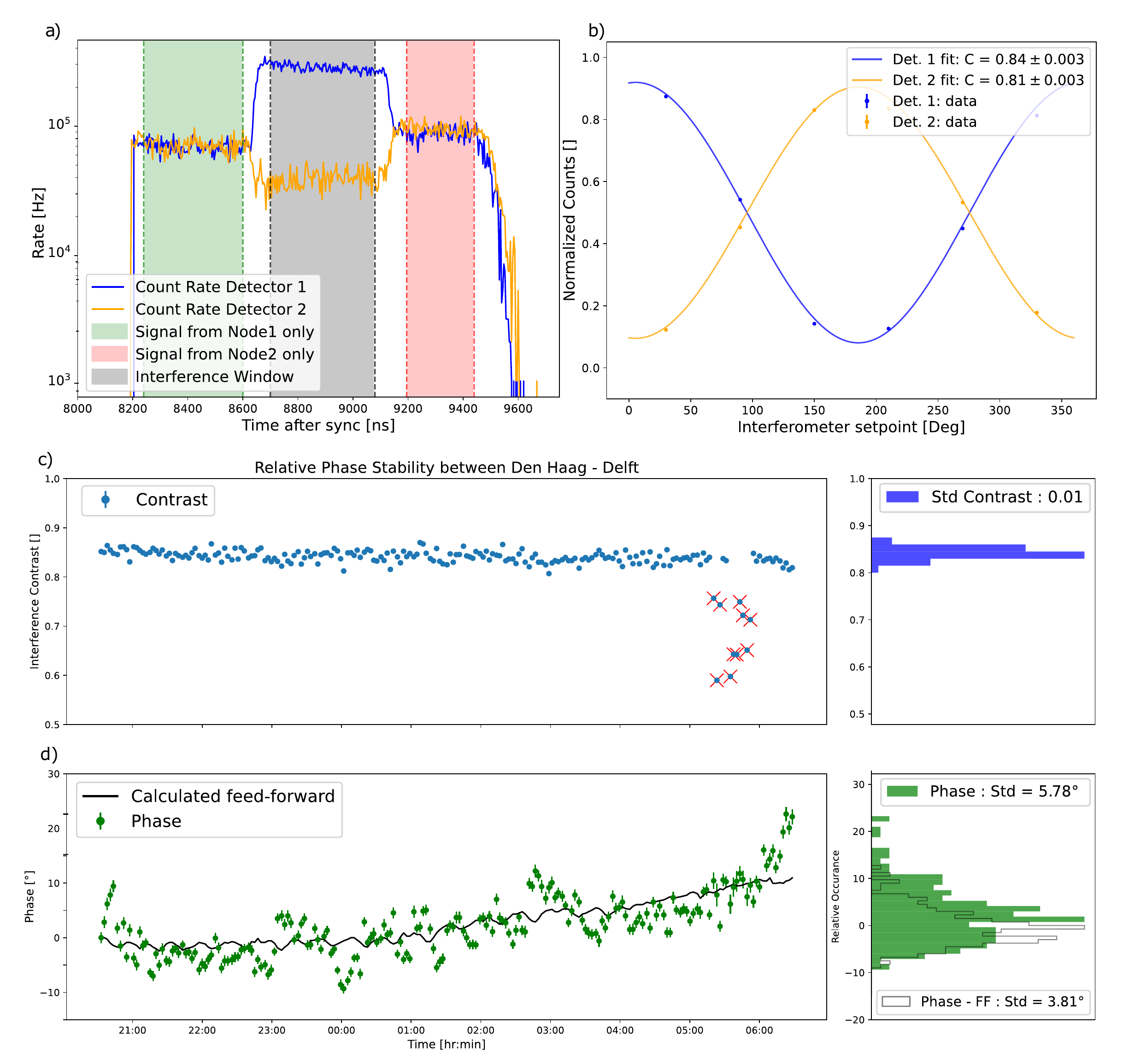}
\caption{\label{fig:totalphaseperformance}\textbf{Performance of the total system performing phase synchronization.}
\textbf{a)} Interference of reflections of the excitation lasers from the diamond chip from, as measured by the SNSPDs in the midpoint. The blue (orange) line indicate the rate of photons incident on the first (second) detector. The shaded indicate regions in time where only light from Node 1 (green) or Node 2 (red) is present show no interference. In the overlapping region (grey), the light interferes and is directed to predominantly one detector, indicating interference of the weak laser fields. \textbf{b)} By sweeping the local oscillator of the global synchronization task in Fig~\ref{fig:controlscheme}c, we change the setpoint of the interference, directing light into one or the other detector, as show by change of preferred direction to detector 1 (blue) or 2 (orange). By correcting the count rates for imbalance, we can fit these oscillations with a single cosine, recovering the relative phase of the excitation lasers of the two nodes. The imperfect contrast of this oscillation represent the residual phase noise in the system at timescales below the measurement time ($\approx$seconds). \textbf{c)} Measurement of the relative optical phase between the paths coming from Delft and The Hague over a timespan of ten hours. We measure a constant average contrast over the full duration of more than 200 measurement, with the exception of a group of 10 measurements. These were the result of one of the excitation lasers lasing multi-mode, as picked up by a separate monitoring scope. \textbf{d} The measured relative phase shows a small drift over this time-frame, without the compensation for phase slips due to large fiber drifts (green, see~\ref{suppsec:defassumptions}). Plotting the calculated feed-forward based on the round-trip time between the two nodes and the midpoint shows the correlation. When taking this feed-forward into accounts, the resulting spread standard deviation of the phase setpoint is below \SI{4}{\degree} over the course 10 hours (white histogram).}
\end{figure*}

We assess the optical performance by using classical light fields reflected off the diamond surface, which should result in homodyne interference, and allows us to access the relative optical phase between the nodes. This light originates in the nodes and travels the same optical path as the ZPL photons (which are coherent with this light), up to a small propagation through the diamond sample. The intensity of this reflected light can be adjusted on the nodes ($>$\SI{1}{\kilo\C\per\second}), allowing for quick integration times and high signal-to-noise ratio in the SNSPDs. We conduct this measurement under identical conditions as an entanglement generation experiment. This means that the powers of the optical fields, both reflected and direct, and their on/off modulation are done in the way one does during entanglement generation. The reflected laser pulses are located at the same location in the sequence as where the optical excitation would be, making the phase measured in this experiment a good metric for the expected performance.

The interference and subsequent measurement by the SNSPDs of this reflected light is show in Fig.~\ref{fig:totalphaseperformance}a, with clear interference shown in the region where the light fields overlap in time. The contrast of this interference can be corrected for imbalance of the incident power, and by sweeping the phase setpoint a full fringe can be taken (Fig.~\ref{fig:totalphaseperformance}b). Fitting these fringes with a single cosine $I = \frac{1 \pm C\cos({\phi + \phi_{0}})}{2}$, we can retrieve the contrast $C$ and setpoint $\phi_0$ of the synchronised system. By assuming that the loss of contrast $C$ is solely due to a normally distributed residual phase error, we calculate its standard deviation from the contrast which is $\sigma_{measured} =$\SI{35.5}{\degree}. This matches well with the expected residual phase noise, which would be the quadratic sum of all five independent synchronization tasks: $\sigma_{total} = \sqrt{2\sigma_{local}^{2} + 2\sigma_{fast}^2 + \sigma_{global}^2} = $\SI{34.9}{\degree}. By repeating this measurement and recording both contrast and setpoint, we can characterize the long-term behaviour of the phase synchronization system (Fig.~\ref{fig:totalphaseperformance}c). We find that the phase synchronization is stable over a time-span of more than ten hours, showing a high contrast. The long term drift (Fig.~\ref{fig:totalphaseperformance}d) of the phase setpoint is almost completely explained by the aforementioned offset in phase due to the combination of the frequency offset and long fiber length variation, which was calculated but not compensated in this test. The resulting phase setpoint distribution is shown in the bottom right of Fig~\ref{fig:totalphaseperformance}d, showing a sharp distribution around a setpoint of \SI{0}{\degree}.

\section{Discussion and Outlook}
We have designed, built and evaluated an extendable phase synchronization scheme to enable entanglement generation between solid-state emitters, compatible with the telecom L band and large node separation. The phase synchronization is achieved between two independent excitation lasers, which are integrated in an NV-center quantum network node, connected over deployed fibers. The introduction of stabilization light allows to generate reliable phase measurements with low shotnoise and enable the ability to guide the stabilization light to an alternative detector. Therefore we can use two distinct feedback loops at the midpoint, to compensate both the high-frequent distortions via a balanced photodetector but also compensate the low-frequent distortions using the SNSPDs, achieving synchronization at the central beamsplitter where the single-photon paths of two distant nodes interfere. The separation of the overall synchronization task and tailoring the control schemes to the specific noise present allow us to realize high-performing synchronization using independent excitation lasers and at a metropolitan scale node separation. By making the synchronization tasks robust against drifts of optical power and compensating for large fiberlength variations, phase stability sufficient for entanglement generation is achieved for more than 10 hours .

Our implementation has a few advantages. First, our method only requires the distribution of a phase reference in the RF domain, e.g. \SI{10}{\mega\hertz}, over the network. This can be done over deployed fiber using the White Rabbit~\cite{Dierikx_whiterabbitprotocol_2016} protocol, and is easily distributed locally via amplified buffers. Second, due to the choice of AOMs as feedback actuators and additional offloading on the pump laser of the QFC, the system has both a fast step response, and the feedback range is only limited by the QFC conversion bandwidth. Additionally, the achieved feedback bandwidth lowers the requirements on the linewidth of our lasers used on the nodes, simplifying their design and avoiding complex optical reference distribution. Furthermore, the design allows the reduction of the optical intensity of signals that could interfere with the entanglement generation, by using heterodyne schemes. This boosts the signal-to-noise ratio of the generated error signal, and at the same time limits the crosstalk with the quantum system and single-photon emission.

This scheme also allows for scaling the number of nodes in the network. The most straightforward way of scaling would be in a star pattern, as all the incoming nodes could be continuously synchronized with the same optical reference. To establish a connection between two nodes, one has to switch the incoming photons to the same beamsplitter. With the appropriate feedback speed one can make the newly switched paths phase-stable before the photons arrive from the remote nodes. A similar approach can be taken on a line-configuration with multiple midpoints if the frequency difference between the optical references used at each midpoint can be kept reasonably small, below the feedback bandwidth of the fast synchronization.

Further improvements can be made for the local and fast subtasks, which we will discuss starting with the local synchronization task. Currently the fully analog and integral nature via feedback on an AOM causes a high gain at low frequencies. This means that it is sensitive to small analog input offsets to this AOM when the errorsignal generation is paused. This results in the reduction of the free-evolution time, the time that the phase stays synchronised without feedback. Using a proportional feedback or a fast digital controller could circumvent this. 
The residual phase noise present in the fast synchronization could be reduced in a number of ways. Excitation lasers with less phase-noise would lead to a direct improvement of the performance. Additionally, more in-depth tuning of the control to the noise-spectrum measured could deal more effectively with the noise. Further investigations into the exact noise spectrum and more complex control techniques could further enhance the performance of the fast synchronization.

As a conclusive demonstration of the system we recently used it to generate heralded entanglement between two NV-centers over deployed metropolitan fiber~\cite{preprint}. Here, the phase synchronization was time-multiplexed with the emission of single-photons by the NV-centers. Being compatible with many different qubit platforms, the architecture presented in this work can form the basis for future developments and implementations generating large-scale entanglement over deployed fiber networks.

\begin{acknowledgments}
We acknowledge funding from the Dutch Research Council (NWO) through the project ``QuTech Part II Applied-oriented research'' (project number 601.QT.001), VICI grant (project no. 680-47-624) and the Zwaartekracht program Quantum Software Consortium (project no. 024.003.037/3368). We further acknowledge financial support from the EU Flagship on Quantum Technologies through the project Quantum Internet Alliance (EU Horizon 2020, grant agreement no. 820445) and from the Dutch Ministry of Economic Affairs and Climate Policy (EZK), as part of the Quantum Delta NL programme, and Holland High Tech through the TKI HTSM (20.0052 PPS) funds.

\textbf{Author contributions}
AJS, JJBB, KLvdE, LvD, EvZ and RH devised the experiment. AJS and JJBB measured and analyzed the experimental data. AJS and JJBB wrote the manuscript. AJS wrote the supplementary information with input from JJBB. RH supervised the experiments. All authors contributed to the manuscript.

\textbf{Data and materials availability.}
The data that support the findings of this study are available upon reasonable request.

\end{acknowledgments}
\newpage

\bibliography{main}

\begin{thebibliography}{10}

\bibitem{kimble_quantum_2008}
H.~J. Kimble, ``The quantum internet,'' {\em Nature}, vol.~453, pp.~1023--1030, (2008).

\bibitem{wehner_quantum_2018}
S.~Wehner, D.~Elkouss, and R.~Hanson, ``Quantum internet: {A} vision for the road ahead,'' {\em Science}, vol.~362, p.~30337383, (2018).

\bibitem{moehring_entanglement_2007}
D.~L. Moehring, P.~Maunz, S.~Olmschenk, K.~C. Younge, D.~N. Matsukevich, L.-M. Duan, and C.~Monroe, ``Entanglement of single-atom quantum bits at a distance,'' {\em Nature}, vol.~449, pp.~68--71, (2007).

\bibitem{ritter_elementary_2012}
S.~Ritter, C.~Nölleke, C.~Hahn, A.~Reiserer, A.~Neuzner, M.~Uphoff, M.~Mücke, E.~Figueroa, J.~Bochmann, and G.~Rempe, ``An elementary quantum network of single atoms in optical cavities,'' {\em Nature}, vol.~484, pp.~195--200, (2012).

\bibitem{hofmann_heralded_2012}
J.~Hofmann, M.~Krug, N.~Ortegel, L.~Gérard, M.~Weber, W.~Rosenfeld, and H.~Weinfurter, ``Heralded {Entanglement} {Between} {Widely} {Separated} {Atoms},'' {\em Science}, vol.~337, pp.~72--75, (2012).

\bibitem{stephenson_high-rate_2020}
L.~J. Stephenson, D.~P. Nadlinger, B.~C. Nichol, S.~An, P.~Drmota, T.~G. Ballance, K.~Thirumalai, J.~F. Goodwin, D.~M. Lucas, and C.~J. Ballance, ``High-{Rate}, {High}-{Fidelity} {Entanglement} of {Qubits} {Across} an {Elementary} {Quantum} {Network},'' {\em Physical Review Letters}, vol.~124, p.~110501, (2020).

\bibitem{Daiss2021}
S.~Daiss, S.~Langenfeld, S.~Welte, E.~Distante, P.~Thomas, L.~Hartung, O.~Morin, and G.~Rempe, ``A quantum-logic gate between distant quantum-network modules,'' {\em Science}, vol.~371, p.~614–617, (2021).

\bibitem{KrutyanskiyEntanglement2023}
V.~Krutyanskiy, M.~Galli, V.~Krcmarsky, S.~Baier, D.~A. Fioretto, Y.~Pu, A.~Mazloom, P.~Sekatski, M.~Canteri, M.~Teller, J.~Schupp, J.~Bate, M.~Meraner, N.~Sangouard, B.~P. Lanyon, and T.~E. Northup, ``Entanglement of trapped-ion qubits separated by 230 meters,'' {\em Phys. Rev. Lett.}, vol.~130, p.~050803, (2023).

\bibitem{Barrett_Kok_2005}
S.~D. Barrett and P.~Kok, ``Efficient high-fidelity quantum computation using matter qubits and linear optics,'' {\em Physical Review A}, vol.~71, p.~060310, (2005).

\bibitem{Hensen_bell_2015}
B.~Hensen, H.~Bernien, A.~E. Dr{\'e}au, A.~Reiserer, N.~Kalb, M.~S. Blok, J.~Ruitenberg, R.~F.~L. Vermeulen, R.~N. Schouten, C.~Abell{\'a}n, W.~Amaya, V.~Pruneri, M.~W. Mitchell, M.~Markham, D.~J. Twitchen, D.~Elkouss, S.~Wehner, T.~H. Taminiau, and R.~Hanson, ``Loophole-free bell inequality violation using electron spins separated by 1.3 kilometres,'' {\em Nature}, vol.~526, pp.~682--686, (2015).

\bibitem{Narla_Shankar_Hatridge_Leghtas_Sliwa_Zalys_Geller_Mundhada_Pfaff_Frunzio_Schoelkopf_etal._2016}
A.~Narla, S.~Shankar, M.~Hatridge, Z.~Leghtas, K.~M. Sliwa, E.~Zalys-Geller, S.~O. Mundhada, W.~Pfaff, L.~Frunzio, R.~J. Schoelkopf, and M.~H. Devoret, ``Robust concurrent remote entanglement between two superconducting qubits,'' {\em Physical Review X}, vol.~6, p.~031036, (2016).

\bibitem{Knaut_SiV_35kmBoston_2023}
C.~M. Knaut, A.~Suleymanzade, Y.-C. Wei, D.~R. Assumpcao, P.-J. Stas, Y.~Q. Huan, B.~Machielse, E.~N. Knall, M.~Sutula, G.~Baranes, N.~Sinclair, C.~De-Eknamkul, D.~S. Levonian, M.~K. Bhaskar, H.~Park, M.~Lončar, and M.~D. Lukin, ``Entanglement of nanophotonic quantum memory nodes in a telecom network,'' {\em Nature}, vol.~629, pp.~573--578, (2024).

\bibitem{Cabrillo_Cirac_García-Fernández_Zoller_1999}
C.~Cabrillo, J.~I. Cirac, P.~García-Fernández, and P.~Zoller, ``Creation of entangled states of distant atoms by interference,'' {\em Physical Review A}, vol.~59, p.~1025–1033, (1999).

\bibitem{Bose_Knight_Plenio_Vedral_1999}
S.~Bose, P.~L. Knight, M.~B. Plenio, and V.~Vedral, ``Proposal for teleportation of an atomic state via cavity decay,'' {\em Physical Review Letters}, vol.~83, p.~5158–5161, (1999).

\bibitem{bernien_heralded_2013}
H.~Bernien, B.~Hensen, W.~Pfaff, G.~Koolstra, M.~S. Blok, L.~Robledo, T.~H. Taminiau, M.~Markham, D.~J. Twitchen, L.~Childress, and R.~Hanson, ``Heralded entanglement between solid-state qubits separated by three metres,'' {\em Nature}, vol.~497, pp.~86--90, (2013).

\bibitem{Slodička_Hétet_Röck_Schindler_Hennrich_Blatt_2013}
L.~Slodička, G.~Hétet, N.~Röck, P.~Schindler, M.~Hennrich, and R.~Blatt, ``Atom-atom entanglement by single-photon detection,'' {\em Physical Review Letters}, vol.~110, p.~083603, (2013).

\bibitem{Delteil_Sun_Gao_Togan_Faelt_Imamoğlu_2016}
A.~Delteil, Z.~Sun, W.-b. Gao, E.~Togan, S.~Faelt, and A.~Imamoğlu, ``Generation of heralded entanglement between distant hole spins,'' {\em Nature Physics}, vol.~12, p.~218–223, (2016).

\bibitem{stockill_phase-tuned_2017}
R.~Stockill, M.~J. Stanley, L.~Huthmacher, E.~Clarke, M.~Hugues, A.~J. Miller, C.~Matthiesen, C.~Le~Gall, and M.~Atatüre, ``Phase-{Tuned} {Entangled} {State} {Generation} between {Distant} {Spin} {Qubits},'' {\em Physical Review Letters}, vol.~119, p.~010503, (2017).

\bibitem{Lago-Rivera_telecommemories_2021}
D.~Lago-Rivera, S.~Grandi, J.~V. Rakonjac, A.~Seri, and H.~de~Riedmatten, ``Telecom-heralded entanglement between multimode solid-state quantum memories,'' {\em Nature}, vol.~594, pp.~37--40, (2021).

\bibitem{LiuTelecomMemories2024}
J.-L. Liu, X.-Y. Luo, Y.~Yu, C.-Y. Wang, B.~Wang, Y.~Hu, J.~Li, M.-Y. Zheng, B.~Yao, Z.~Yan, D.~Teng, J.-W. Jiang, X.-B. Liu, X.-P. Xie, J.~Zhang, Q.-H. Mao, X.~Jiang, Q.~Zhang, X.-H. Bao, and J.-W. Pan, ``Creation of memory–memory entanglement in a metropolitan quantum network,'' {\em Nature}, vol.~629, p.~579–585, (2024).

\bibitem{RuskucREIentanglement2024}
A.~Ruskuc, C.-J. Wu, E.~Green, S.~L.~N. Hermans, J.~Choi, and A.~Faraon, ``Scalable multipartite entanglement of remote rare-earth ion qubits,'' 2024.

\bibitem{Humphreys_deterministicdelivery_2018}
P.~C. Humphreys, N.~Kalb, J.~P.~J. Morits, R.~N. Schouten, R.~F.~L. Vermeulen, D.~J. Twitchen, M.~Markham, and R.~Hanson, ``Deterministic delivery of remote entanglement on a quantum network,'' {\em Nature}, vol.~558, pp.~268--273, (2018).

\bibitem{Pfaff2014}
W.~Pfaff, B.~J. Hensen, H.~Bernien, S.~B. van Dam, M.~S. Blok, T.~H. Taminiau, M.~J. Tiggelman, R.~N. Schouten, M.~Markham, D.~J. Twitchen, and R.~Hanson, ``Unconditional quantum teleportation between distant solid-state quantum bits,'' {\em Science}, vol.~345, p.~532–535, (2014).

\bibitem{pompili_multinodenetwork_2021}
M.~Pompili, S.~L.~N. Hermans, S.~Baier, H.~K.~C. Beukers, P.~C. Humphreys, R.~N. Schouten, R.~F.~L. Vermeulen, M.~J. Tiggelman, L.~dos Santos~Martins, B.~Dirkse, S.~Wehner, and R.~Hanson, ``Realization of a multinode quantum network of remote solid-state qubits,'' {\em Science}, vol.~372, no.~6539, pp.~259--264, 2021.

\bibitem{Hermans_Pompili_Beukers_Baier_Borregaard_Hanson_2022}
S.~L.~N. Hermans, M.~Pompili, H.~K.~C. Beukers, S.~Baier, J.~Borregaard, and R.~Hanson, ``Qubit teleportation between non-neighbouring nodes in a quantum network,'' {\em Nature}, vol.~605, p.~663–668, (2022).

\bibitem{preprint}
A.~J. Stolk, K.~L. van~der Enden, M.-C. Slater, I.~t. Raa-Derckx, P.~Botma, J.~van Rantwijk, B.~Biemond, R.~A.~J. Hagen, R.~W. Herfst, W.~D. Koek, A.~J.~H. Meskers, R.~Vollmer, E.~J. van Zwet, M.~Markham, A.~M. Edmonds, J.~F. Geus, F.~Elsen, B.~Jungbluth, C.~Haefner, C.~Tresp, J.~Stuhler, S.~Ritter, and R.~Hanson, ``Metropolitan-scale heralded entanglement of solid-state qubits,'' 2024.

\bibitem{Dréau_Tchebotareva_Mahdaoui_Bonato_Hanson_2018}
A.~Dréau, A.~Tchebotareva, A.~E. Mahdaoui, C.~Bonato, and R.~Hanson, ``Quantum frequency conversion of single photons from a nitrogen-vacancy center in diamond to telecommunication wavelengths,'' {\em Physical Review Applied}, vol.~9, p.~064031, (2018).

\bibitem{Geus_Elsen_Nyga_Stolk_vanderEnden_vanZwet_Haefner_Hanson_Jungbluth_2023}
J.~Geus, F.~Elsen, S.~Nyga, A.~J. Stolk, K.~L. van~der Enden, E.~J. van Zwet, C.~Haefner, R.~Hanson, and B.~Jungbluth, ``Low-noise short-wavelength pumped frequency down-conversion for quantum frequency converters,'' {\em Optica Open}, (2023).

\bibitem{Tchebotareva_TelecomSpinPhotonEntanglement_2019}
A.~Tchebotareva, S.~L.~N. Hermans, P.~C. Humphreys, D.~Voigt, P.~J. Harmsma, L.~K. Cheng, A.~L. Verlaan, N.~Dijkhuizen, W.~de~Jong, A.~Dr\'eau, and R.~Hanson, ``Entanglement between a diamond spin qubit and a photonic time-bin qubit at telecom wavelength,'' {\em Phys. Rev. Lett.}, vol.~123, p.~063601, (2019).

\bibitem{Stolk_VanDerEnden_Roehsner_Teepe_Faes_Bradley_Cadot_VanRantwijk_TeRaa_Hagen_etal_2022}
A.~J. Stolk, K.~L. van~der Enden, M.-C. Roehsner, A.~Teepe, S.~O.~J. Faes, C.~E. Bradley, S.~Cadot, J.~Van~Rantwijk, I.~Te~Raa, R.~A.~J. Hagen, A.~L. Verlaan, J.~J.~B. Biemond, A.~Khorev, R.~Vollmer, M.~Markham, A.~M. Edmonds, J.~P.~J. Morits, T.~H. Taminiau, E.~J. Van~Zwet, and R.~Hanson, ``Telecom-band quantum interference of frequency-converted photons from remote detuned nv centers,'' {\em PRX Quantum}, vol.~3, p.~020359, (2022).

\bibitem{Chen2020DRS}
J.-P. Chen, C.~Zhang, Y.~Liu, C.~Jiang, W.~Zhang, X.-L. Hu, J.-Y. Guan, Z.-W. Yu, H.~Xu, J.~Lin, M.-J. Li, H.~Chen, H.~Li, L.~You, Z.~Wang, X.-B. Wang, Q.~Zhang, and J.-W. Pan, ``Sending-or-not-sending with independent lasers: Secure twin-field quantum key distribution over 509 km,'' {\em Phys. Rev. Lett.}, vol.~124, p.~070501, (2020).

\bibitem{patent_sync}
J.~J.~B. Biemond, A.~J. Stolk, K.~L. van~der Enden, E.~J. van Zwet, and A.~J.~H. Meskers, ``Controlled synchronisation at bandwidths beyond the limit of direct synchronisation error measurement.'' Patent pending, 2023.

\bibitem{patent_chromaticdiff}
J.~J.~B. Biemond, A.~J.~H. Meskers, W.~D. Koek, A.~J. Stolk, K.~L. van~der Enden, and E.~J. van Zwet, ``Estimation method of phase difference between two optical channels induced by optical latency variations.'' Patent pending, 2023.

\bibitem{Dierikx_whiterabbitprotocol_2016}
E.~F. Dierikx, A.~E. Wallin, T.~Fordell, J.~Myyry, P.~Koponen, M.~Merimaa, T.~J. Pinkert, J.~C.~J. Koelemeij, H.~Z. Peek, and R.~Smets, ``White rabbit precision time protocol on long-distance fiber links,'' {\em IEEE Transactions on Ultrasonics, Ferroelectrics, and Frequency Control}, vol.~63, no.~7, pp.~945--952, 2016.

\bibitem{mat_kor_23}
A.~Matsuki, H.~Kori, and R.~Kopayashi, ``An extended {Hilbert} transform method for reconstructing the phase from an oscillatory signal,'' {\em Scientific Reports}, vol.~13, no.~3535, 2023.

\bibitem{Welch1967}
P.~Welch, ``The use of fast fourier transform for the estimation of power spectra: A method based on time averaging over short, modified periodograms,'' {\em IEEE Transactions on Audio and Electroacoustics}, vol.~15, p.~70–73, (1967).

\bibitem{Stroud_1971}
C.~R. Stroud, ``Quantum-electrodynamic treatment of spontaneous emission in the presence of an applied field,'' {\em Physical Review A}, vol.~3, p.~1044–1052, (1971).

\end{thebibliography}
\bibliographystyle{my_plain}
\onecolumngrid

\clearpage
\begin{center}
\textbf{\large Supplementary Materials}
\end{center}
\makeatletter
\renewcommand{\theequation}{S\arabic{equation}}
\renewcommand{\thefigure}{S\arabic{figure}}
\renewcommand{\thetable}{S\arabic{table}}
\renewcommand{\thesection}{S-\Roman{section}}
\makeatother
\setcounter{equation}{0}
\setcounter{figure}{0}
\setcounter{table}{0}
\setcounter{section}{0}
\subsection{System identification details}
\begin{figure}[h]
\includegraphics[width=0.8\columnwidth]{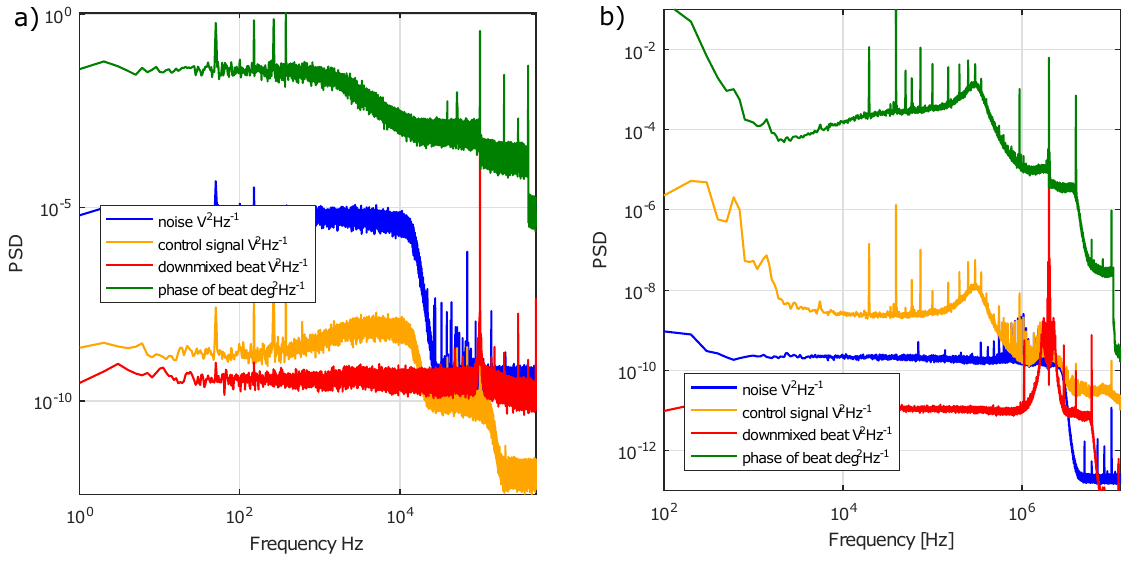}
\caption{\label{fig:psdscombined}\textbf{System identification of synchronization tasks}.a) Power Spectral Density of noise injected before the controller $\eta_\text{idL1}$, control output signal $u_{L1}$, measured beat signal $m_{L1}$ (analoguely downmixed to \SI{100}{\kilo\hertz}), and phase error on this signal, respectively, for the local synchronization task. b) Power Spectral Density of noise injected before the controller $\eta_\text{idf1}$, control output signal $u1$, measured beat signal $m1$ (analoguely downmixed to \SI{2}{\mega\hertz}), and phase error on this signal, respectively, for the fast synchronization task. 
}
\end{figure}

We injected additional noise at the location shown in Fig.~\ref{fig:controlscheme} while the control loop is active and measured the signals $u_{L1}$ and the demodulated output $d_{L1}$ of the local synchronization loop. Throughout the paper, these measured interference powers $x(t_i), i=1,\ldots, N,$  are analysed by subsequently taking the {Hilbert} transform to compute the phase of the resulting analytical signal~\cite{mat_kor_23}. Additionally we can plot the spectral content measured during the identification experiments. This is done with Welch's averaged periodogram method~\cite{Welch1967}, and is shown in Fig.~\ref{fig:psdscombined}. We repeat this analysis for the fast synchronization loop, injecting noise, monitoring the signals and demodulated output, and perform the analysis. The spectral content of the measurements are show in Fig.~\ref{fig:psdscombined}, with the analysis in the main text.

For the global synchronization task injecting noise posed a challenge, and we therefore only measured the residual phase error under open/closed loop conditions. 

\captionsetup[figure]{justification=raggedright}

\section{Theoretical description of synchronization result}\label{suppsec:theodescription}
Here we give a theoretical description of the phase synchronization scheme, and determine the conditions that must hold for the phase synchronization to succeed. We start off by describing some definitions and assumptions underlying the analysis and provide a schematic that includes all relevant fields and optical paths. This allows us to define the central stabilization task we want to achieve, and how we achieve it by breaking up the task into smaller synchronization tasks. 

\subsection{Definitions and assumptions}
\label{suppsec:defassumptions}
We make the following definitions and assumptions in the derivation of the phase conditions:
\begin{itemize}
    \item In propagating the fields we treat them all  as monochromatic plane waves that propagate along the optical axis, the $\widehat{z}$ direction. We can therefore write the complex wavefunction of a field with frequency $f_i$ in a medium with refractive index $n_{l}$ depending on time $t$ and space $z$ as  \begin{equation}
        U(z, t) =  Ae^{j\phi_i}= Ae^{j(\omega_{i} t - k_{i} z)}
    \end{equation}
    with $A$ being a constant, $\omega_{i} = 2\pi f_i$ and $k_{i} = \frac{n\ \omega_{i}}{c}$ with $c$ the speed of light in vacuum and $n$ the refactive index. We ignore all effects of a varying spatial intensity distribution or curvature of the wavefront in free space. 
    \item We treat reflection and transmission through the Ultra-narrow filter as having no impact on the phase, an assumption that is true if the central frequency is kept stable, and the reflection is far away from the resonance of the filter.
    \item We assume that the change in refractive index for small frequency differences ($<$\SI{1}{\giga\hertz}) or temperature variations are negligible.
    \item All clocks (RF sources as references in the synchronization) used in the scheme are coherent with each other. This is realized by providing a \SI{10}{\mega\hertz} External Reference to each of the signal generators, and finding clock settings that minimize any non-idealities in the signal generation of the devices used.
\end{itemize}
We make extensive use of heterodyne interference, which is the interference of two fields of different frequencies. When measuring the intensity of this field with a photodiode, the combined field intensity is described as
\begin{equation}
    I = |U_1 + U_2|^2 = |U_1|^2 + |U_2|^2 + U_1^*U_2 + U_1U_2^* 
\end{equation}
Filling in two plane waves of equal amplitude $\sqrt{I_0}$ at the start of the lasers, but different frequencies and initial phase, we get
\begin{equation}\label{eq:heterodynebeat}
I(t) = 2I_0 + 2I_0\cos\big({(\omega_2 - \omega_1)t +\frac{n}{c}(\omega_1 D_1 - \omega_2 D_2) + (\theta_2 - \theta_1)\big)}, 
\end{equation}
where THz-frequencies are dropped.
The intensity is varying in time with the difference of the two frequencies of the original waves (called a beat), with change of the distances $D_1$ and $D_2$, and with a phase given by the initial phases of the two fields. This allows us to use the photodiode to generate an error signal that contains both the difference frequency $\omega_1 - \omega_2 = \Delta\omega$ and the relative phase of the two optical waves. By choosing the frequency difference of the two optical fields in the RF regime ($<$\SI{1}{\giga\hertz}), we can use RF-generators, mixers, amplifiers and filters to stabilize this errorsignal. This synchronizes the two fields, and adjusts for frequency variations, drifts of the distances to the point of interference, and phase jumps due to the linewidth of the laser. This is the underlying principle to stabilize the relative optical phase of two optical fields, at a certain location in space.

It is useful to give a proper definition of what we mean when we say two phases are synchronized. When two optical fields in a task are \textit{synchronized}, it means that at a specific point in space $z_{s}$, the two fields $U_{1}(z_{s}, t)$ and $ U_{2}(z_{s}, t)$ have a phase relation that can be written as 
    \begin{equation}
    \big(Arg(U_{1}(z_{s}, t)) - Arg(U_{2}(z_{s}, t))\big) - \omega_{clock}t  \equiv \eta(t)
    \end{equation}
with $\omega_{clock}$ the clock frequency of the RF-source used in the signal processing and $\eta(t)-\mu \ll \pi$ the residual phase error and $\mu$ a constant. For sampling frequencies much smaller than the feedback bandwidth (and therefore slower than phase residuals), these samples are independent of their evaluation time $t_i,\i=0,1,2,...$, and thus follow a probability distribution with constant parameters in time, i.e.
\begin{equation}
\label{eq:sampling}
\eta(t_i)\in\mathcal{N}(\mu,\sigma),\ i=0,1,2,... ,
\end{equation} where, for small angles, $\mathcal{N}$\ is a normal distributed variable with mean $\mu$ and standard deviation $\sigma$. Experimentally we can choose $\mu \in [0, 2\pi]$ by changing the clock setpoint used for the stabilization, and $\sigma \ll \pi$ gives the performance of the synchronization: smaller $\sigma$ indicates better synchronization performance. For the special case where $\omega_{clock} = 0$, or when one has access to a source coherent with the clock, one can sample the distribution $\mathcal{N}$. For entanglement generation experiments this conditions holds, where the sampling occurs at the measurement of a single-photon every couple of seconds.

\begin{figure*}[h]
\centering
\includegraphics[width=0.3\linewidth]{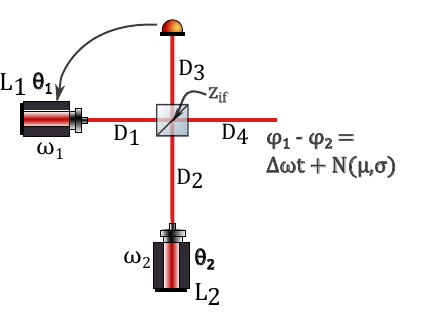}
\caption{\label{suppfig:simple_phase_synch}\textbf{Phase synchronization.} An example of two independent sources $1$ and $2$, being synchronized by using an interference measurement on one output $D_3$ of a beamsplitter, traveling further via $D_4$. The interference location $z_{if}$ is the point on the beamsplitter where the beams meet. Depending on the setpoint and performance of the synchronization task, the relative phase $\phi_1 - \phi_2$ can be written as a time dependent term, plus a normal distribution $\mathcal{N}$ with mean $\mu$ and standard deviation $\sigma$ that do not depend on time.}
\end{figure*}

\subsection{Errors due to large length variations}\label{subsec:phaseslipsub}
Suppose we have two coherent, co-propagating fields $U_1$ and $U_2$ with different frequencies $\omega_1, \omega_2$, on a long fiber with length L. The first field is (perfectly) synchronized at the end of the fiber $z = L$ with a phase actuator that works \textbf{on both} $U_1$ and $U_2$, with the phase setpoint $\theta_{Ref}:=\frac{nL\omega_1}{c}$. The phase that the field $U_2$ has at $z = L$ is than given as \begin{equation} \left.\varphi_2\right|_{z=L}=
    \frac{nL\omega_2}{c} = \theta_{Ref} + \frac{n\ L(\omega_2 - \omega_1)}{c} = \theta_{Ref} +  \frac{n\ L(\omega_2 - \omega_1)}{c},
\end{equation} which is at a constant offset of $\theta_{ref}$. This means the synchronization task of $U_1$ is also synchronizing $U_2$, it is no longer varying in time, albeit at a constant offset. If now the fiber expands to length $L + \Delta L$, the synchronization tasks aims for phase $\left.\varphi_1\right|_{z=L+\Delta L}\to \theta_{Ref}$. Hence,
 \begin{align}
    \left.\varphi_2\right|_{z=L+\Delta L}&=\frac{n\ (L + \Delta L)\omega_2}{c} = \frac{n\ (L+\Delta L) \omega_1}{c} + \frac{n\ (L + \Delta L)(\omega_2 - \omega_1)}{c} =\theta_{Ref} +  \frac{n\ L(\omega_2 - \omega_1)}{c} +   \frac{n\ \Delta L(\omega_2 - \omega_1)}{c} \\
    &=    \frac{n\ L\omega_2}{c} +\theta_{err}(\Delta L),
\end{align} with $\theta_{err}(\Delta L)=\frac{n\ \Delta L(\omega_2 - \omega_1)}{c}$ an additional term, indicating the phase of $U_2$ has moved with respect to its original position at $L$, and is no longer synchronized. therefore we have to be careful in exchanging phase terms of fields that are co-propagating and synchronized over long fibers, but at slightly different frequencies. Using the numbers in our system, $\Delta \omega = 2 \pi \times$\SI{400}{\mega\hertz}, and a $\Delta L$ of \SI{2}{\centi\meter} expansion of the fiber would result in a $\theta_{err}(0.02)$ of \SI{10}{\degree} of phase error. Given the low thermal expansion coefficient of silica of $\frac{dL}{dT} = 5.5 \times 10^-7$\SI{}{\meter\per\kelvin}, this effect only happens when large fiber lengths and temperature drifts are involved. In the main text, Fig. \ref{fig:totalphaseperformance} shows a measurement of this effect, where the value calculated using our accurate timing hardware matches the value measured by the phase interference well.

\begin{figure*}[h]
\includegraphics[width=0.9\linewidth]{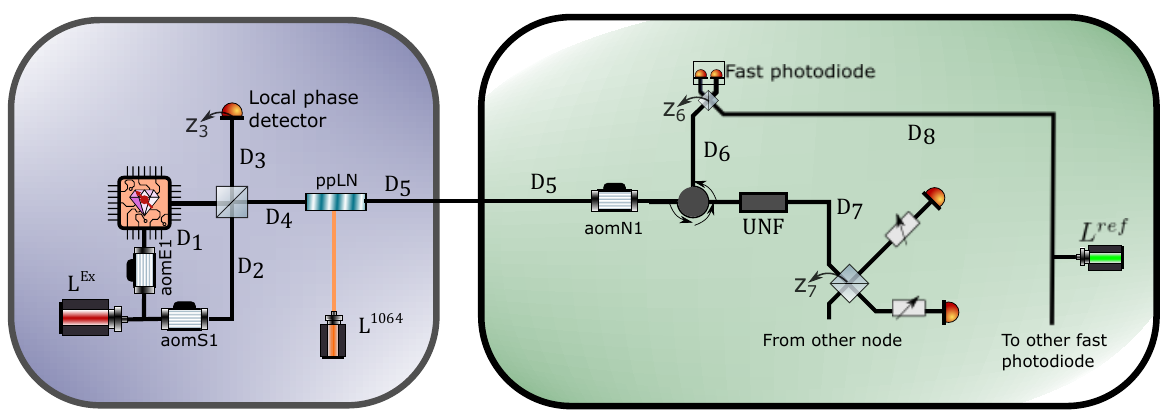}
\caption{\label{suppfig:opticalpaths}\textbf{Schematic showing the division of optical paths in the system, used for deriving the synchronization tasks.} The main synchronization task, and the subdivision of local, fast and global synchronization tasks is given in the main text Fig. \ref{fig:opticsphaselock}}
\end{figure*}

For a full overview of all the optical fields used, we refer to Table \ref{tab:phasestabparam} that gives the typical central frequency, linewidth and powers. In Fig.~\ref{suppfig:opticalpaths} we show the division of the connection of one node to the midpoint into separate optical paths, used in the derivation of the phase synchronization requirements. These sections ($D1$, $D4$, $D5$ and $D7$) form a continuous path between the excitation laser $L^{Ex}$ and the central beamsplitter in the midpoint where the relevant interference takes place, plus additional sections that are needed to describe the individual synchronization tasks ($D2$, $D3$, $D6$ and $D8$). A short description, typical distances and extra information is given in Table \ref{tab:opticalpaths}.

\begin{table*}[!h]
\centering
\begin{tabular}{|l|l|l|l|l|l|l|l|}
\hline
Optical field             & NV emission    & Excitation & Reflections    & Stabilization & Leakage                 & QFC Pump & Reference    \\ \hline\hline
Typ. Wavelength [GHz]       & 470450       & 470450    & 470450        & 470450.4  & 470450.4            & 281759  & 188691     \\ \hline
Typ. Linewidth [kHz] & 12.3e3         & $\sim$10   & $\sim$10       & $\sim$10      & $\sim$10                & \textless{}20  & \textless{}0.1 \\ \hline
Power in Local PD           & Single photons & $\sim$5uW  & $\sim$1nW      & $\sim$5uW     & -                       & -        & -            \\ \hline
Power in Midpoint        & Single photons       & -          & \textless{}1pW & $\sim$30nW   & Single Photons (100kHz) & -        & $\sim$5mW    \\ \hline
\end{tabular}
    \caption{\label{tab:phasestabparam}Overview of optical fields.}
\end{table*}
\newpage

\begin{table*}[]
\centering
\begin{tabular}{|p{0.05\linewidth}|p{0.45\linewidth}|p{.05\linewidth}|p{.4\linewidth}|}
\hline
\hline
Section  & Description                                                                                            & Length [m] & Dominant noise source                                  \\ \hline
$D_{1}$ & Excitation path, from laser, via AOM,  objective and diamond chip to local central beamsplitter       & 10             & Objective stage/vibrations and laser phase noise                            \\ \hline
$D_{2}$ & Local stabilization light path, from laser, via AOM, free-space optics, to local central beamsplitter & 10             & Fibre optics/vibrations                               \\ \hline
$D_{3}$ & Local central beamsplitter to local phase projection (Polarizing BS)                                             & 2              & Long term optics drift/birefringent optics            \\ \hline
$D_{4}$ & Local central beamsplitter, via free-space optics, to QFC crystal                                      & 2              & -                                                     \\ \hline
$D_{5}$ & QFC crystal, via deployed fibers, through EPC, AOM to Ultra-narrow filter                              & $>$10e3      & Thermal expansion/environmental vibrations \\ \hline
$D_{6}$ &Backwards reflection off UNF, to BS at fast detector                                                              & 2              & Vibrations/loose fibers                               \\ \hline
$D_{7}$ & Transmission through UNF, towards central beamsplitter in midpoint                                     & 2              & Vibrations/loose fibers                               \\ \hline
$D_{8}$ & Reference laser, via fibers, to BS at fast detector                                                        & 2              & Vibrations/loose fibers                               \\ \hline
\end{tabular}
\caption{\label{tab:opticalpaths}Overview and description of optical paths.}
\end{table*}

\subsection{Central synchronization task}

We define the central synchronization task as achieving a stable phase between the photon emission on each node, measured at the location of the central beamsplitter in the midpoint, denoted $z_7$ in Fig.~\ref{suppfig:opticalpaths}. This is the central requirement for entanglement generation via the single-click protocol, as discussed in the main text.
To simplify things, we will first derive the stability of the relative phase between the two excitation lasers on each node, as measured in the central beamsplitter in the midpoint, and then argue that the single photon emission is coherent with this field (see Sec. \ref{sec:conclusion}).

The time-dependent relative phase at the central beamsplitter can be written as
\begin{equation}\label{eq:centralsynctask}
\eta(t) = \phi(t, z_7)^{1} - \phi(t, z_7)^{2}   
\end{equation}
where $\phi(t, z_7)^i$ is the phase of the field coming from node $i$ at the location of the central beamsplitter $z_7$. When all the synchronization tasks are active, the error $\eta(t)$ will be small ($\ll\pi$), and under probabilistic entanglement generation samples this time-dependent phase is sampled, and the errors form a distribution $\mathcal{N}_{Tot}$.

We can write $\phi(t, z_7)$ by following the fields through the setup as shown in the main text Fig.~\ref{fig:opticsphaselock}. It is given by 
\begin{equation}\label{eq:centraltaskinput}
   \phi(t, z_7) = \theta_{Ex} + \frac{n \omega_{Ex}}{c}(D_1 + D_4) - \theta_{1064} + \\ \frac{n' \omega_{Ex}'}{c}(D_5 + D_7) + \omega_{Ex}'t,
\end{equation}
where $\omega_{Ex}$ the frequency of the excitation laser,  $\theta_{Ex/1064}$ are the phases of the excitation and QFC pump laser respectively. The $'$ denotes the fact that the field has been converted, meaning $\omega_{Ex}' = \omega_{Ex} - \omega_{1064}$. For the other node to the midpoint we can write an identical equation, containing the same but uncorrelated terms. In order to synchronize these two fields, we need to get rid of the terms that are varying rapidly in time (see Tab. \ref{tab:opticalpaths}).
The next step is to fill in the synchronization conditions guaranteed by the synchronization tasks realized in the setup. 

\subsection{Local synchronization task}
The local synchronization task synchronizes the relative phase of the light coming from two paths consisting of $D_1+D_3$, and $D_2+D_3$, both starting at the excitation laser and ending at detector $z_3$. In the path $D_2+D_3$, the AOM shifts the light in frequency with \SI{400}{\mega\hertz}$\pm$\SI{750}{\hertz}. Due to the two paths being different polarization states of the light, the actual interference happens after projection into the same state by a polarizer, and subsequent detection by the detector after travelling $D_3$, which we denote as $z_3$, see Fig.~\ref{fig:opticsphaselock}b and  Fig.~\ref{suppfig:opticalpaths}. The synchronization error is given by the equation \begin{equation*} \eta(t)_{loc} = (\phi^{A}(t, z_3) - \phi^{B}(t, z_3)) - \omega_{loc}t,\end{equation*} and can be written out as 
\begin{equation}\label{eq:loc}
  \omega_{Loc}t + \eta_{loc}(t) = \Big(\theta_{Ex} +  \frac{n \omega_{Ex}}{c}(D_1+D_3) + \omega_{Ex}t \Big)- \Big(\theta_{S} +  \frac{n \omega_{S}}{c}(D_2+D_3) + \omega_{S}t \Big)
\end{equation}
with $\theta_{Ex}$ the phase due to the finite linewidth of the Excitation laser, and $\omega_{S}$ the frequency of the stabilization light and $\omega_{loc} = \omega_{Ex} - \omega_{S}$ the frequency and $\theta_{loc}$ the phase of the clock used as reference. By ensuring that $D_1 - D_2$ is well within the coherence length of the excitation laser, we can write $\theta_{Ex} = \theta_{S}$. The remaining terms are slowly varying in time with respect to the feedback bandwidth, as shown in the main text \ref{fig:localmerged}. The performance parameter $\sigma_{loc}$ is also given there.
We can also write \ref{eq:loc} in a different form that makes it easier to use in future derivation as 
\begin{equation}\label{eq:locrewrite}
    \theta_{Ex} +  \frac{n \omega_{Ex}}{c}D_1 = \theta_{S} +  \frac{n \omega_{S}}{c}D_2 +  \frac{n (\omega_{S} - \omega_{Ex})}{c}D_3 + \eta_{loc}(t),
\end{equation}

\subsection{Fast synchronization task}
The next synchronization task we describe is the fast synchronization task. This consist of again two arms, $A$ and $B$, that meet at the point $z_{6}$. Path $A$ is from the excitation laser, via the stabilization split-off ($D_2$), through the QFC ($D_{4}$), over the long deployed fiber ($D_{5}$), through the AOM at the midpoint, reflected by the UNF ($D_{6}$) to reach $z_6$. Path $B$ is from the Reference laser directly to $z_6$ (via $D_8$), see Fig.~\ref{fig:opticsphaselock}b and Fig.~\ref{suppfig:opticalpaths}. We can again define the synchronization error for this task as 
\begin{equation*}
    \eta(t)_{fast} = (\phi^{A}(t, z_6) - \phi^{B}(t, z_6)) - \omega_{fast}(t),
\end{equation*} and filling in the terms we find

\begin{multline}\label{eq:fastsynch}
    \omega_{fast}t + \eta_{fast}(t) = \\ \Big(\theta_{S} + \frac{n \omega_{S}}{c}(D_2 + D_4) - \theta_{1064} + \frac{n'\omega_{S}'}{c}(D_5 + D_6) + \omega_{S}'t \Big) - \Big( \theta_{Ref} + \frac{n\ D_8\ \omega_{Ref}}{c} + \omega_{Ref}t \Big)
\end{multline}
Where $\omega_{fast} = \omega_{S}' - \omega_{Ref}$ is the frequency and $ \theta_{loc}$ the phase of the reference clock used in the fast lock (see Fig.~\ref{fig:controlscheme}b). Again, in order for this synchronization to work, we need to have fast enough feedback with respect to the noise present. Because this equation contains the three phase term due to the linewidth of the lasers ($\theta_{Ex/1064/Ref}$), of which $\theta_{Ex}$ is by far the dominant (see Table~\ref{tab:phasestabparam}). For long fibers, the term containing the long fibers ($D_5$) can also induce fast phase fluctuations. We show that we can ensure these conditions in the main text, Fig.~\ref{fig:fast1merged}, and give the performance parameter $\sigma_{fast}$.
We can rewrite \ref{eq:fastsynch} to a more useful form as
\begin{multline}\label{eq:fastrewrite}
    \theta_{S} + \frac{n \omega_{S}}{c}(D_2 + D_4) - \theta_{1064} + \frac{n'\omega_{S}'}{c}D_5 + \omega_{S}'t \\ = -\frac{n'\omega_{S}'}{c}D_6 + \omega_{fast}t + \eta_{fast}(t) + \theta_{Ref} + \frac{n\ D_8\ \omega_{Ref}}{c}
\end{multline}

\subsection{Global synchronization task}
The final synchronization task is the global synchronization, which is the final task that closes the feedback system between the two distant nodes. It takes as input two arms, $A$ and $B$, that interfere at the point $z_7$, the central beamsplitter. The fields that are interfering is stabilization light that is leaking \textit{through} the UNF in each arm. This is light that came from the same location as in the fast synchronization task \ref{eq:fastsynch}, but is now travelling towards the SNSPDs. We can write the optical phase coming from node $A/B$ as

\begin{equation}\label{eq:globalinput}
\phi^{A/B}(t, z_7) = \theta_{Ex} + \frac{n \omega_{S}^{A/B}}{c}(D_{2}^{A/B} + D_{4}^{A/B}) - \theta_{1064} + \frac{n'\ \omega_{S}^{'A/B}}{c}(D_{5}^{A/B} + D_{7}^{A/B}) + \omega_{S}^{'A/B}t
\end{equation}

We described in the main text that the global synchronization task only needs a small bandwidth (\SI{1000}{\hertz}) to be realized. However, equation \ref{eq:globalinput} contains many fast terms, such as the laser linewidth of the Excitation laser. Therefore the global synchronization can only be realized once the fast synchronization condition is met. This becomes apparent when we filling in \ref{eq:fastrewrite} in \ref{eq:globalinput}, giving:
\begin{equation}
\phi^{A/B}(t, z_7) = \eta_{fast}(t) + \theta_{Ref} +
\frac{n\ D_{8}^{A/B}\ \omega_{Ref}}{c} + \frac{n'\ \omega_{S}^{'A/B}}{c}(D_{7}^{A/B} - D_{6}^{A/B}) + \omega_{S}^{'A/B}t
\end{equation}and using that to write down synchronization task error for the global lock:
\begin{equation}
    \eta(t)_{glob} = \big(\phi^{A}(t, z_7) - \phi^{B}(t, z_7)\big) - \omega_{glob}(t) - \theta_{glob}(t)
\end{equation}
which, when grouping similar terms pairwise, becomes
\begin{align*}
     \eta(t)_{glob} = &\eta_{fast}^{A}(t) - \eta_{fast}^{B}(t) + \stepcounter{equation}\tag{\theequation} \label{eq:globalsync} \\
     &\theta_{Ref}^{A}- \theta_{Ref}^{B} + \frac{n\ (D_8^{A} - D_8^{B})\ \omega_{Ref}}{c} +\\
     &\frac{n'\ \omega_{S}'^{A}}{c}(D_7^{A} - D_6^{A}) - \frac{n'\ \omega_{S}'^{B}}{c} (D_7^{B} - D_6^{B}) + \\
     &(\omega_{S}'^{A} - \omega_{S}'^{B})t -\theta_{glob}(t)\\
\end{align*}
where the super-script$^{A/B}$ denotes the field coming from node $1/2$ and $\omega_{glob} = (\omega_{S}'^{A} - \omega_{S}'^{B})$ the frequency and $\theta_{glob}$ the phase of the clock used as reference in the global synchronization (see Fig.~\ref{fig:controlscheme}c). We will go over this equation term by term. 
The first line of the equation are the residual terms from the fast synchronization tasks $\eta_{fast}^{A/B}(t)$, which are the residuals of the fast lock, and therefore too fast for the global synchronization to provide any feedback on. That is why they also appear on the other side of the equation, as they remain an error source in the system. The second line is the phase contribution due to the linewidth of the reference laser and optical path of the delivery of the laser to the fast photo-diodes. By using the same length fibers ($D^{A}_8 \sim D^{B}_8$) we can minimize both the effect of fiber drifts, as well as the effect of the Reference laser linewidth. 
The third line of equation \ref{eq:globalsync} shows two terms that are dependent on the distance from the UNF to the two points (fast photo-diode $z_6$ and central beamsplitter $z_7$) of interference. These are more difficult to keep equal, and is best practice to keep these fibers in the same optical rack such that they share vibrations/temperature fluctuations.
We can also identify the synchronization error of the global synchronization \textit{by itself} by excluding the fast errors and writing as 
\begin{equation}\label{eq:globalrewrite}
    \eta_{glob}(t) = \theta_{Ref}^{A}- \theta_{Ref}^{B} + \frac{n\ (D_8^{A} - D_8^{B})\ \omega_{Ref}}{c} + \frac{n'\ \omega_{S}'^{A}}{c}(D_7^{A} - D_6^{A}) - \frac{n'\ \omega_{S}'^{B}}{c} (D_7^{B} - D_6^{B}) - \omega_{glob}t
\end{equation}
Now that we have discussed all three different synchronization tasks and described their synchronization task error, we can go back go to \ref{eq:centralsynctask} and fill in the synchronization conditions.

\subsection{Substitution Local synchronization}
We start with equation \ref{eq:centraltaskinput} and fill in the local synchronization condition in the form of Eq. \ref{eq:locrewrite} to get
\begin{equation}\label{eq:localfilledin}
       \phi(t, z_7) = \theta_{S} +  \frac{n \omega_{S}}{c}D_2 + \eta_{loc}(t)+ \frac{n (\omega_{S} - \omega_{Ex})}{c}D_3+\frac{n \omega_{Ex}}{c}D_4 - \theta_{1064} + \frac{n' \omega_{Ex}'}{c}(D_5 + D_7) + \omega_{Ex}'t
\end{equation}
where the term $\frac{n (\omega_{S} - \omega_{Ex})}{c}D_3 = \theta_{err}(\Delta D_3)$ is due to the fact that the interference for the local synchronization happens at a distance $D_3$ away from where the stabilization light $\omega_S$ is split off (local beamsplitter). The phase is therefore sensitive to expansion/contraction of $D_3$, albeit with only the frequency difference of $\omega_{S} - \omega_{Ex}$. Therefore drifts of $D_3$ are minimized by making the distance short and housed in a temperature stabilized environment and can be considered constant.

\subsection{Substitution fast synchronization}
In order to fill in the fast synchronization task into eq. \ref{eq:localfilledin}, we need to substitute all the terms containing $\omega_{Ex}$ with $\omega_{S}$. These fields are co-propagating from the local beamsplitter towards the QFC ($D_4$) and onward. Because the local synchronization ensure their mutual coherence at the local beamsplitter, and $\omega_{S}$ is stabilized in the midpoint at the fast photodiode, we can follow the substitution method as outlined in Section~\ref{subsec:phaseslipsub}, taking into account the expected fiber drifts. The terms containing $\omega_{Ex}$ contain the distances $D_4, D_5$ and $D_7$. We remark that the drifts in $\theta(\Delta D_4)$ and $\theta(\Delta D_7)$ are negligible when the (fiber) lengths are short($\approx$\SI{10}{\meter}) and will be considered constant and left out, however $\theta(\Delta D_5)$ can be significant due to the length of $D_5$ (see Table~\ref{tab:opticalpaths}).
Completing the substitution gives us:
\begin{multline}\label{eq:stabsubstituted}
\phi(t, z_7) = \theta_{S} +  \frac{n \omega_{S}}{c}(D_2+D_4)+\theta_{err}(\Delta D_3, \omega_{loc}) + \eta_{loc}(t)-\\\theta_{1064}  + \frac{n' \omega_{S}'}{c}(D_5 + D_7) + \theta_{err}(\Delta D_5, \omega_{loc}) + (\omega_{S}' + \omega_{loc})t
\end{multline}
where we have added the error term $\theta_{err}(\Delta D_5, \omega_{loc})$ accordingly. We can now recognize the terms of the fast synchronization task, and filling in the form of eq. \ref{eq:fastrewrite} we arrive at the expression

\begin{equation}
\label{eq:fastfilledin}
       \phi(t, z_7) = \eta_{loc}(t) + \eta_{fast}(t) + \theta_{err}(\Delta D_5, \omega_{loc}) + \\ \theta_{Ref} + \frac{n\ \omega_{Ref}}{c}D_8 + \frac{n' \omega_{S}'}{c}(D_7 - D_6) + (\omega_{fast} + \omega_{loc})t
\end{equation}
\subsection{Substitution global synchronization}
We can now fill in eq. \ref{eq:fastfilledin} into \ref{eq:centralsynctask}, and use the global synchronization condition (the last four terms of this expression are precisely one half of the global synchronization condition \ref{eq:globalrewrite}), to simplify it into its final form: 
\begin{align}
\label{eq:finerr}
    \eta(t) = &\eta_{loc}^{A}(t) + \eta_{fast}^{A}(t) - \eta_{loc}^{B}(t)) + \eta_{fast}^{B}(t) + \\ &\theta_{err}^{A}(\Delta D_5, \omega_{loc})-\theta_{err}^{B}(\Delta D_5, \omega_{loc}) + \\ &\Omega_{tot}t + \eta_{glob}(t) 
\end{align}
The first line is the performance of the local and fast synchronization task $\eta_{loc/fast}^{A/B}(t)$ of the two nodes, the second line the error due to the expansion/contraction of the long fibers $\theta_{err}^{A/B}(D_5)$. The last line is the global synchronization task performance $\eta_{glob}(t)$, and a time dependent term $\Omega_{tot}\ t$, where $\Omega_{tot} = (\omega_{fast}^A + \omega_{loc}^A) - (\omega_{fast}^B + \omega_{loc}^B)$. All the other terms in this equation are either not dependent on time, or we have discussed ways to minimize their effects over longer time duration. The only thing left is then to choose the right frequencies of all the synchronization tasks such that the time dependence of the phase is removed, e.g. $\Omega_{tot} = 0$, with only small variations of $\eta(t)$ remaining in the system.

\subsection{Selection of clock frequencies}
As described in the main text, the condition $\Omega_{tot} = 0$ is not the only requirement for the frequencies at which the synchronization tasks operate. Due to shot-noise limitations of the SNSPDs, $\omega_{S} '^A - \omega_S'^B = \omega_{glob}$ can realistically not exceed \SI{10}{\kilo\hertz}, and in order for the feedback bandwidth of the fast synchronization to be fast enough, a high $\omega_{fast}$ is needed.
The additional requirement of $\omega_{Ex}'^{A} = \omega_{Ex}'^{B}$ to generate indistinguishable photons, while the natural frequency $\omega_{Ex}^{A} - \omega_{Ex}^{B}>$\SI{1}{\giga\hertz} can be far detuned, adds to the complexity. The flexibility in choosing $\omega_{1064}$ on each of the nodes separately deals with this requirement, and shifts both the excitation and stabilization light equally. Considering other experimental details outside the scope of this work regarding the temperature stabilization of the FBGs, we arrive on the choice of frequencies as presented in Table \ref{tab:synctasks}. Setting $\omega_{loc}^{A/B} = 400e6 \mp 750$Hz and $\omega_{fast}^{A} = 215001500$Hz and $\omega_{fast}^{B} = 215000000$Hz. With these values, the remaining frequency difference between the excitation lasers is:  \begin{equation}
    \Omega_{tot} = (215001500 + 399999250) - (215000000 + 400000750) = 0
\end{equation} 
Additionally, this gives us $\omega_{glob} = 1500$Hz, which lies well within the bandwidth of the SNSPDs.

\subsection{Conclusion}
\label{sec:conclusion}
We can now return to the central synchronization condition \ref{eq:centralsynctask} and make the following claim. Given that the conditions \begin{enumerate}\label{it:conditions}
    \item Both nodes have the local phase synchronization active  (\ref{eq:loc}),
    \item The midpoint synchronizes the Stabilization light to the Reference laser (\ref{eq:fastsynch}),
    \item The Stabilization light of the nodes is synchronized to each other(\ref{eq:globalsync}), 
    \item The known error $\theta_{err}^{A/B}(\Delta D_5)$ is calculated and compensated by adjusting the phase of the clock used for reference in the global synchronizaiton.
    \item The phase is sampled at a frequency below the frequency of the noise (\ref{eq:sampling}),
    \item The measured residual phase noise in each of the synchronization tasks can be considered independent,
\end{enumerate}
hold we can consider the excitation lasers to be synchronized. Condition six is valid due to the fact that the local synchronization task feedback is done on the excitation laser AOM, and therefore not affected by the synchronization in the midpoint. Additionally the residual phase noise of the fast synchronization is practically zero below the bandwidth of the global synchronization task, see Fig. ~\ref{fig:fast1merged} and  ~\ref{fig:globalpsd}, making them independent.
Given these conditions, then the sampling of the error $\eta(t)$ from Eq. \ref{eq:finerr} of the phase between the excitation lasers of each node arriving at the central beamsplitter in the midpoint can be written as:\begin{gather}
    \mathcal{N}_{tot}(\mu_{tot}, \sigma_{tot}) + \theta_{offset}\ \\
    \mu_{tot} = \mu^{A}_{loc} + \mu^B_{loc} + \mu^A_{fast} + \mu^B_{fast} +\mu_{glob} \\
    \sigma_{tot} = \sqrt{(\sigma^{A}_{loc})^2 + (\sigma^B_{loc})^2 + (\sigma^A_{fast})^2 + (\sigma^B_{fast})^2 +(\sigma_{glob})^2}
\end{gather}
where $\mathcal{N}_{tot}(\mu_{tot}, \sigma_{tot})$ is the distribution of the total phase error, and $\mathcal{N}_{i}^{A/B}(\mu_{i}, \sigma_{i})$ the mean and standard deviation of resulting distributions of the sampling of the individual synchronization task errors. The value $\theta_{offset}$ is a term that contains all the constant phase terms in this derivation. This is the central result of the derivation, and provides the full description of the total synchronization. This result is also experimentally verified in Fig.  \ref{fig:totalphaseperformance}, by measuring the interference of the excitation lasers at the central beamsplitter. The measurement involves the changing of the phase of the RF-clock used for the global synchronization, therefore changing $\mu_{global}$. The measured phase of the oscillation of Fig.  \ref{fig:totalphaseperformance}b is the true 'phase setpoint' of the optical interference of the excitation lasers. The relation between $\mu_{global}$ and this phase setpoint is difficult to calculate due to the constant terms neglected in this derivation, currently denoted as $\theta_{offset}$. It can however be measured relatively quickly and consistently, as Fig.~\ref{fig:totalphaseperformance}c shows over the duration of \SI{10}{\hour}, showing that its long-term drift is dominated by the terms mentioned in Section~\ref{it:conditions}4.

\section{Relation between optical phase and entangled state phase}
The main result of this work shows the stabilization of the optical phase between two remote nodes over telecom fiber. The light with which we show this stability is reflected of the diamond chip, and follows the same optical path towards the midpoint and central beamsplitter. However, during the entanglement generation, the Excitation light is actually: \begin{enumerate}
    \item propagating $\approx$\SI{10}{\micro\meter} into the diamond,
    \item excitation the Nitrogen Vacancy-center resonantly and subsequent emission of a single photon through spontaneous emission,
    \item that propagates $\approx$\SI{10}{\micro\meter} back out of the diamond.
\end{enumerate}
All steps add an additional phase to the optical field that is now a single photon. These factors introduce fixed offsets between the phase investigated in this work, and need to be calibrated using a separate experiment. therefore, a stable optical phase is necessary but not sufficient to generate entanglement between two distant NV-center electron spins. Any drifts in these parameters would change the entangled state phase, and therefore the correlations between the spin states.  
These processes can be considered constant in time given the following conditions:
\begin{enumerate}
    \item The path taken towards the NV center is constant in time
    \item The path taken from the NV center towards the collection is constant in time and a single spatial mode with well-defined phase.
    \item The spontaneous emission process is coherent with the exciting field
\end{enumerate}
Both (1) and (2) are realized by making sure the relative position of the microscope objective used to address the NV center and the diamond sample is stationary during the entanglement generation. Furthermore, if the objective is moved during the operation, a check of the entangled state phase has to be done again. The relative microscope and sample location is one of the limiting factors in keeping a constant entangled state phase. The spontaneous emission is found to be coherent with the excitation field \cite{Stroud_1971}, and, because of the short $\approx$\SI{1}{\nano\second} excitation pulse used, even in the presence of small detunings with the emitter.

\newpage

\end{document}